# Atmospheres of Solar System Moons and Pluto[1]

**Xi Zhang[2]**
Department of Earth and Planetary Sciences, University of California Santa Cruz,
Santa Cruz, CA, 95064, United States

## Synopsis

The atmospheres within our Solar System can be categorized into four distinct climate regimes: "terrestrial", "Jovian", "condensable", and "exosphere". Beyond the three terrestrial planets (excluding Mercury) and the four giant planets, collisional atmospheres are also found on smaller celestial bodies such as Jupiter's moon Io, Saturn's moon Titan, Neptune's moon Triton, and Pluto. This article reviews the key characteristics of these atmospheres and the underlying physical and chemical processes that govern them. I focus on their thermal structures, chemical constituents, wind patterns, and the origins and losses of the atmospheres, and highlight the critical roles of surface ices and liquids, atmospheric hazes, and the space environments of their host planets in shaping these atmospheres. I dedicated this article to Prof. Zuo Xiao (1936-2024) at Peking University.

## Key Points

- Io is a volcanic world with a widespread but partially collapsed atmosphere, primarily composed of sulfur dioxide ($SO_2$).
- Titan is the only extra-terrestrial world known to have surface liquids interacting with its thick atmosphere.
- The atmospheres of Io, Triton, and Pluto are thin, significantly influenced by sublimation and condensation of surface ices.
- The atmospheres of Titan, Triton, and Pluto share similar compositions of nitrogen ($N_2$), methane ($CH_4$), and carbon monoxide (CO), as well as photochemically produced hydrocarbons, nitriles, and abundant hazes particles.
- The atmospheres of Io, Titan, and Triton are also largely affected by their host planets' space environment.

---

[1] Chapter in *Encyclopedia of Atmospheric Sciences, Third Edition.*

[2] xiz@ucsc.edu

## 1. Introduction

In the Solar System, atmospheres are usually found on bodies with sufficient gravity to retain them. The size of the object and the heliocentric distance are crucial factors in determining the presence of an atmosphere. If the temperature is high, atmospheres can quickly escape; for example, Mercury is too hot on the dayside, leading to rapid atmospheric loss. Likewise, the Earth's Moon, the big asteroid Ceres, the three Galilean moons of Jupiter (Ganymede, Callisto, and Europa), and Saturn's moon Enceladus only have exospheres, where molecules and atoms are not collisional and can freely escape to space. The innermost Galilean moon, Io, is an exception because its atmosphere is continuously replenished by volcanic plumes due to tidal heating from Jupiter. If the temperature is too low, an atmosphere can also collapse; for example, Eris, a Kuiper Belt object roughly the size of Pluto, might produce a collisional atmosphere when it approaches the Sun, but its atmosphere has yet to be detected. To date, collisional atmospheres have been confirmed on eleven bodies in the Solar System: three terrestrial planets (Venus, Earth, and Mars), four giant planets (Jupiter, Saturn, Uranus, and Neptune), three moons (Io, Titan, and Triton), and Pluto.

The atmospheres of Io, Titan, Triton, and Pluto are the focus of this article, with their basic parameters shown in Table 1. Although these bodies are much smaller than Earth (20-40% Earth's radius), their atmospheres are highly active and exhibit a wide variety of weather behaviors (Figure 1). Over 400 active volcanoes continually resurface Io and sustain its thin, localized atmosphere through volcanic plumes. The sulfur-rich atmosphere of Io escapes, creating extended neutral clouds and a corona around the moon. Titan's surface pressure is 50% higher than Earth's, with a similar $N_2$-dominated atmosphere that interacts with surface seas and lakes to drive an active methane-based hydrological cycle. The organic chemistry in Titan's atmosphere produces abundant hydrocarbon and nitrile aerosols, giving the moon its characteristic yellowish color. Although the atmospheres of Triton and Pluto also share similar compositions ($N_2$/$CH_4$/CO) with Titan and are also rich in hydrocarbons and aerosols, their atmospheres are much thinner (around 1 Pa at the surface). These atmospheres are thermodynamically coupled with surface volatile ices and drive the volatile transport across the globe. The active plumes and large cloud formations on Triton indicating strong geological activities in the icy world. All three moons—Io, Titan, and Triton—have a close relationship with their host planets, where magnetospheric particle precipitation significantly influences their upper atmospheres. The Jupiter-Io interaction, through an electric current, produces intense auroral glows at Jupiter's poles as a footprint of Io. While Pluto lacks a host planet, its escaping atmosphere might extend far enough to produce reddish deposits in the polar regions of its moon, Charon.

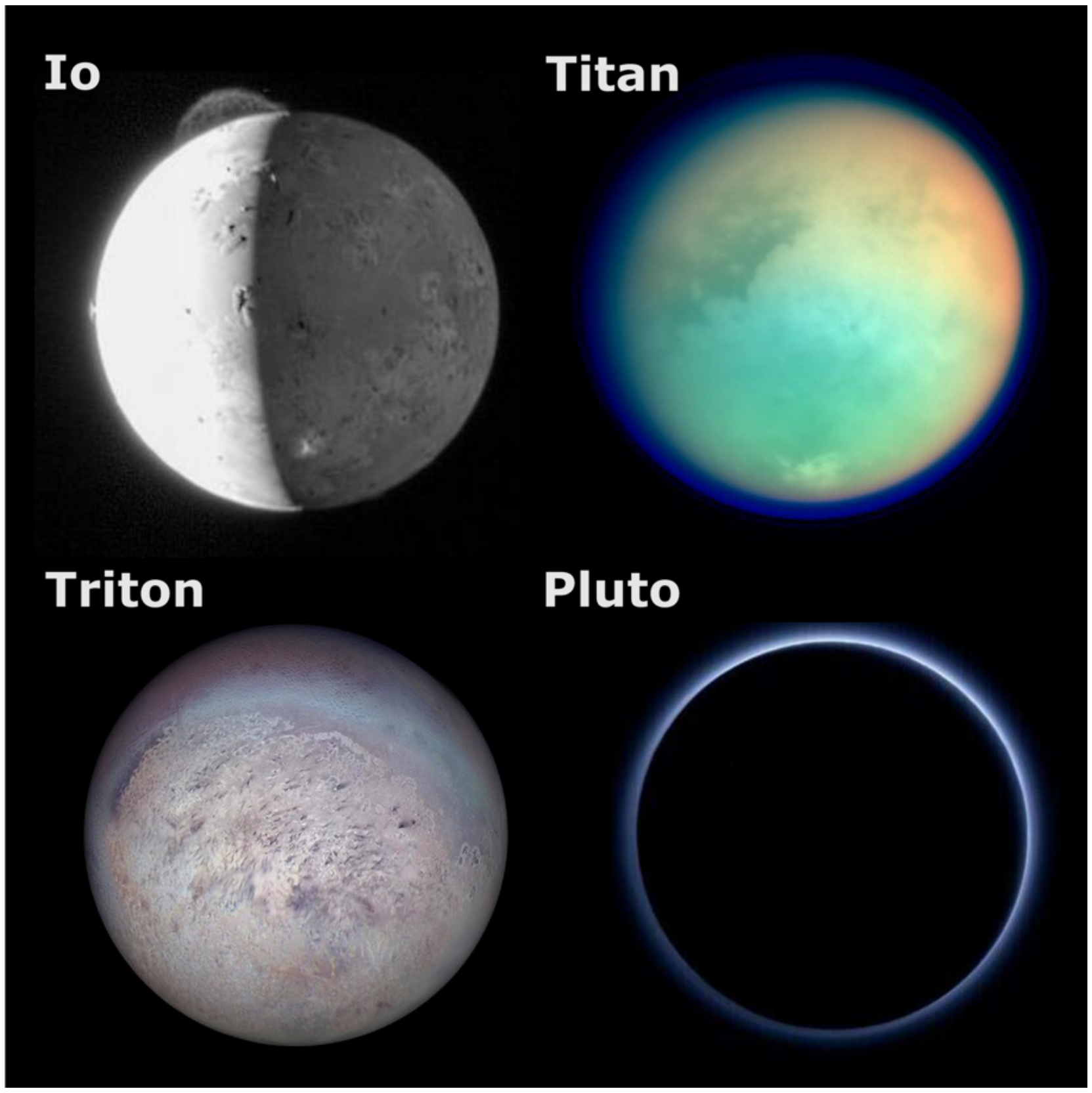

***Figure 1.*** *A family portrait of small bodies in the Solar System with appreciable atmospheres. Image credits from NASA: Io (New Horizons image PIA09248), Titan (Cassini image PIA06139), Triton (Voyager 2 image by NASA/JPL), Pluto (New Horizons image PIA19964).*

***Table 1.*** *Comparison of the Planetary and Atmospheric Properties of Io, Titan, Triton, Pluto, and Earth, based on de Pater et al. (2023), Strobel and Cui (2014), and Gladstone and Young (2019).*

| **Property/body** | **Io** | **Titan** | **Triton** | **Pluto** | **Earth** |
|---|---|---|---|---|---|
| **Mass ($10^{22}$ kg)** | 8.933 | 13.457 | 2.147 | 13.03 | 597.217 |
| **Mean Radius (km)** | 1821.3 | 2575.0 | 1352.6 | 1188.3 | 6371.0 |
| **Distance to the sun (AU)** | 5.20 | 9.54 | 30.07 | 39.48 | 1 |
| **Orbital Eccentricity** | 0.048 | 0.056 | 0.009 | 0.25 | 0.017 |
| **Orbital Period (year)** | 11.86 | 29.42 | 163.72 | 247.7 | 1 |
| **Rotation Period (day)** | 1.77 | 15.95 | 5.88 | 6.39 | 1 |
| **Obliquity (°)** | 3.12 | 26.73 | 129.6 | 119.6 | 23.45 |
| **Surface Gravity (m $s^{-2}$)** | 1.80 | 1.35 | 0.78 | 0.62 | 9.81 |
| **Surface Pressure (Pa)** | (4-40) $x10^{-4}$ | $1.5x10^5$ | 1.2-1.4 | 1.0-1.3 | $1.01x10^5$ |
| **Surface Temperature (K)** | 85-140 | 94 | 38 | 39-54 | 288 |
| **Atmospheric Composition** | ~90% ($SO_2$), ~10% (SO and O) | 95% ($N_2$), <5% ($CH_4$), 0.1% ($H_2$), 0.004% (CO), 0.003% (Ar) | 99% ($N_2$), 0.02% ($CH_4$), 0.01% (CO) | 99% ($N_2$), 0.5% ($CH_4$), 0.05% (CO) | 78% ($N_2$), 21% ($O_2$), <4% ($H_2O$) |
| **Aerosol Mole Fraction** | - | 1 x $10^{-8}$ | 2 x $10^{-7}$ | 4 x $10^{-7}$ | $10^{-9}$–$10^{-7}$ |
| **Exosphere Temperature (K)** | - | 170 | 95 | 70 | 700-1500 |
| **Escape Rate ($s^{-1}$)** | $3x10^{28}$ (S and O) | $9x10^{28}$ ($H_2$), $10^{27}$ (H), Possibly $2x10^{27}$ ($CH_4$) | $10^{24}$ ($N_2$), $10^{26}$ (H) | $5x10^{25}$ ($CH_4$), $10^{23}$ ($N_2$) | $10^{27}$ (H), $8x10^{24}$ (He) |

The atmosphere of Titan was discovered in the early 1940s, Io's in the late 1970s, and those of Triton and Pluto in the late 1980s. Since their discovery, ground-based telescopes have continuously monitored these atmospheres. All these moons have been visited by space probes: the Galileo mission studied Io, the Cassini mission observed Titan (with the Huygens lander descending through Titan's atmosphere), the Voyager spacecraft flew by Triton, and the New Horizons mission recently visited Pluto. These space missions provided invaluable close-up information for atmospheric characterization, while ground-based telescopes have offered complementary long-term tracking of their climate changes over time. This article summarizes the current knowledge of their atmospheres from both observational and theoretical perspectives, following the order of Io, Titan, Pluto, and Triton. Triton is discussed last because we have the least knowledge about its atmosphere, and placing Triton in the context of Titan and Pluto sheds light on how $N_2$/$CH_4$/CO atmospheres function in different parameter spaces.

## 2. Io

As the innermost Galilean satellite, Io stands out as the most active volcanic world in the Solar System. Tidal interactions between Io and Jupiter generate substantial internal heat, producing an average surface heat flux of about 2-3 W $m^{-2}$—significantly higher than Earth's 0.1 W $m^{-2}$. Approximately 54% of Io's total heat flow is attributed to volcanic power output from hot spots. Consequently, Io's surface geology is primarily shaped by intense volcanic and tectonic activities. The vigorous volcanic activities also produce a tenuous yet collisional atmosphere on Io.

The potential for an atmosphere on Io was first suggested during the Pioneer 10 flyby, which detected the ionosphere. Subsequent observations by the Voyager spacecraft in 1979 confirmed volcanic plumes above Pele and Loki, and detected $SO_2$ gas emanating from Loki. We now understand that Io's atmosphere is predominantly composed of $SO_2$, which can exist in a solid state under Io's surface conditions. Photodissociation of $SO_2$ can produce sulfur monoxide (SO) and atomic oxygen (O), both of which were observed in the atmosphere, although SO may also be emitted directly from volcanic eruptions. Additional compounds such as atomic chlorine (Cl), sodium chloride (NaCl), and potassium chloride (KCl) have been detected, likely originating from volcanic activity and/or possibly from surface sputtering at high latitudes. Sulfur ($S_2$), detected once, is thought to arise from volcanic activity. The molecular composition observed in Io's atmosphere aligns with the expected gas emissions from terrestrial volcanoes. Nevertheless, several common volcanic species found on Earth, such as hydrogen sulfide ($H_2S$), water ($H_2O$), and carbon compounds like carbon monoxide (CO), carbon dioxide ($CO_2$), and methane ($CH_4$), are conspicuously absent from Io's atmosphere.

Early observations indicated that Io's local surface pressure ranged from approximately 4 to 40 nanobars, correlating to column densities between $2 \times 10^{17}$ and $2 \times 10^{18}$ $cm^{-2}$. Although Io's atmosphere is believed to be widespread, observations show large spatial variability in density and pressure, with localized density increases over active plumes, longitudinal $SO_2$ density variations, and a latitudinal decrease in gas pressure, particularly beyond 45 degrees latitude. For example, larger column densities of $SO_2$ are consistently found on the anti-Jovian hemisphere (about $10^{17}$ cm$^{-2}$) compared to the sub-Jovian hemisphere (around $10^{16}$ cm$^{-2}$).

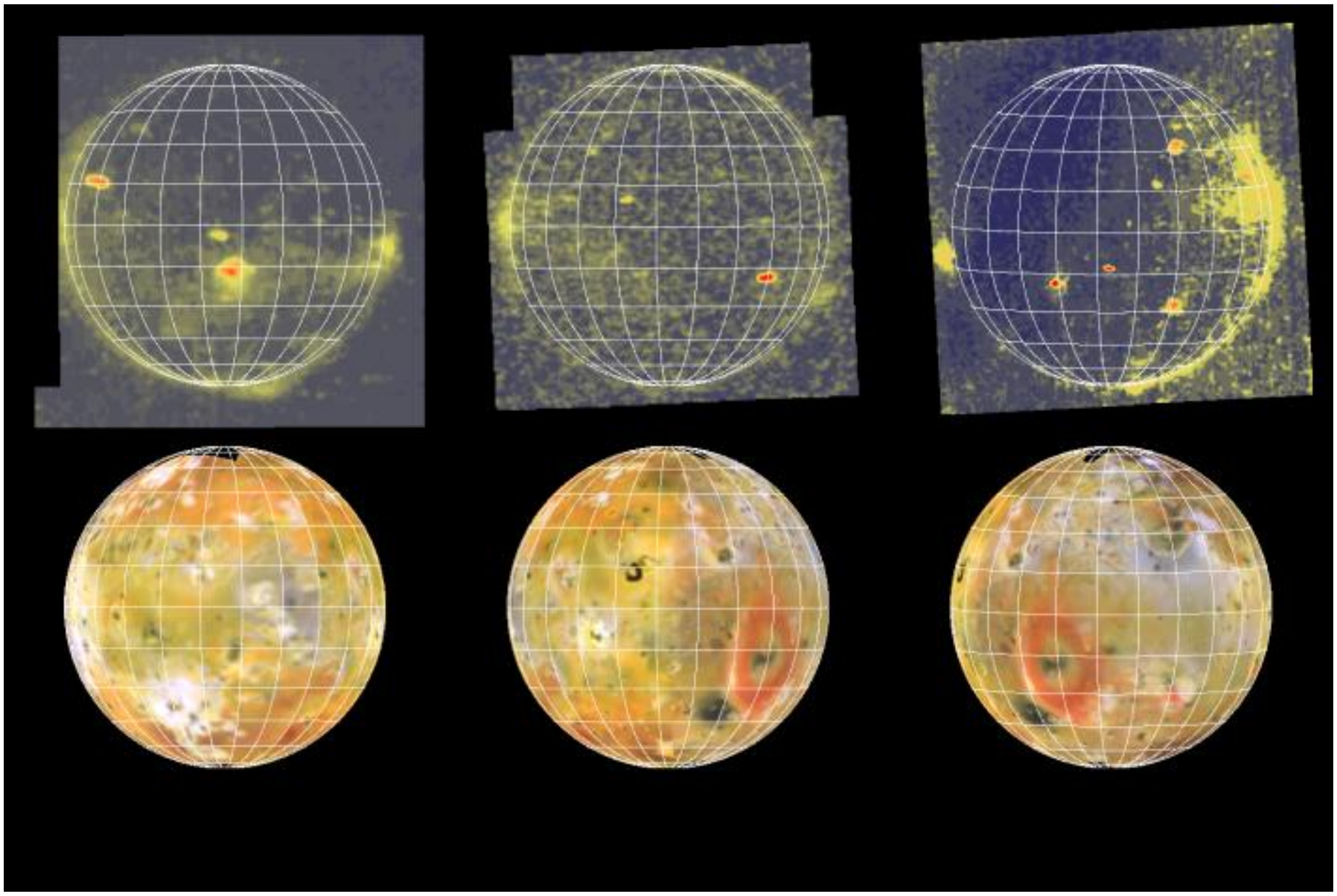

***Figure 2.*** *(top) Io's images during eclipse showing volcanic hot spots and airglow associated with volcanic plumes. Brightness varies from red (high) to dark blue (low). Small red ovals indicate thermal emission from volcanic hot spots above 700 K, while diffuse greenish areas near the limb suggest auroral and airglow emissions from neutral oxygen or sulfur. (bottom) Corresponding images in reflected light. North is to the top. Image credit: NASA's Galileo mission (PIA00739).*

Io's atmosphere has a complex thermal structure, profoundly influenced by the sublimation of $SO_2$, coupled with considerable heating from solar absorption and radiative cooling mechanisms. Non-local thermodynamical equilibrium (non-LTE) effects significantly impact cooling efficiency, with radiative cooling from the vibrational bands leading to a temperature decrease in the lower atmospheric layers. Although observational data vary, they generally suggest atmospheric temperatures ranging from 150 K to 320 K. Infrared observations have identified mean temperatures around 170 K, dipping as low as 108 K on the anti-Jovian hemisphere. Models that incorporate Joule and plasma heating indicate dramatic temperature increases, reaching up to about 1800 K at higher altitudes.

One of the unique and extreme behaviors of Io's atmospheric variability is its response to eclipses (Figure 2). As Io enters Jupiter's shadow, both atmospheric and surface temperatures drop rapidly, leading to the swift condensation of $SO_2$ within minutes. Consequently, Io's atmosphere may undergo partial or complete collapse during eclipse ingress and then rapidly reform upon egress. This phenomenon is supported by the observed "post-eclipse brightening" of Io, which is blanketed with frost immediately after the eclipse. The brightness typically returns after about 10-20 minutes, likely due to the sublimation of the frost. Volcanic activity may help prevent the complete collapse of the $SO_2$ atmosphere. The presence of non-condensable gases, such as SO, may also create a surface layer that hinders the rapid collapse of the condensable component. However, concentrations of SO also show significant decreases during eclipses, possibly due to interactions with the surface.

Io's atmospheric dynamics are greatly influenced by large temperature and density variations, which give rise to complex wind patterns. The extreme spatial contrasts, characterized by virtually no atmosphere in the polar regions or on the nightside, drive supersonic winds. Early models suggested that winds generally flow from the warmer subsolar point towards the colder nightside and high latitudes, propelled by horizontal pressure gradients. However, observations reveal strong horizontal winds moving in the prograde direction, challenging these early simple models. While wind patterns near the terminator may still flow from day to night, the comprehensive picture of wind behavior also includes the influences of sublimation-driven flow, volcanic activity, and thermal effects such as the day-to-night atmospheric collapse. Simulations indicate that different wind patterns may exist for $SO_2$ and SO as the terminator is approached and on the nightside. The most advanced simulations have demonstrated that Io's atmosphere may exhibit complex wind features, such as a prograde equatorial jet, cyclonic and anticyclonic winds at higher latitudes, and interactions with the surface that lead to wind dissipation as the atmosphere becomes thinner.

The atmospheric dynamics are also greatly shaped by localized disturbances caused by volcanic plumes. These plumes, a mixture of dust and gas with dust constituting a significant but variable fraction of the total mass, are visible in sunlight due to the scattering of light off fine dust particles. The plumes create colorful surface deposits, including bright rings from "Pele-type" eruptions, and are composed of $SO_2$ frost, various sulfur allotropes, and sulfur polymorphs. Plumes are categorized into two types based on their height and deposits: high-altitude "Pele-type" plumes, which reach over 400 km and are marked by red rings, and lower-altitude "Prometheus-type" plumes, which rise less than 100 km. Large eruptions, such as those seen with Pele and Prometheus, can persist for decades, while other plumes exhibit unsteady dynamics with fluctuations occurring over the span of minutes. The plume gas could rapidly expand and flow over the top of the sublimation atmosphere, forming a volcanic atmosphere that is potentially warmer and chemically distinct, overlaying and exerting pressure on a sublimation atmosphere which might partially condense back onto the surface.

The atmospheric variability of Io is likely shaped by an interplay between sublimation, volcanic activity, and sputtering. Sublimation and condensation between the surface $SO_2$ frost and vapor are important, leading to atmospheric density peaking at perihelion and diminishing at aphelion. This sublimation-sensitive atmosphere exhibits swift changes, particularly observable during eclipses, though its diurnal variability is less pronounced, likely due to the frost's high thermal inertia. Volcanism plays a key role in shaping Io's atmosphere through the expulsion of gases like $SO_2$, SO, S, NaCl, and KCl, particularly from plumes that reach great altitudes. Volcanic activity, alongside sublimation, may dominate the atmospheric composition in certain regions, contributing to the overall atmospheric pressure and initiating complex wind patterns. Sputtering, induced by Jupiter's magnetospheric particles, is most pronounced during the moon's eclipse or nighttime or at higher latitudes where other processes are minimal, resulting in a darker surface and reduced $SO_2$ at these latitudes.

The redistribution of NaCl and KCl suggests that while volcanic activity may initially release these compounds, sputtering likely contributes to their spread across Io's atmosphere. Io's atmosphere has a short residence time of about 10 days because it continuously loses material at a rate of about 1 ton per second due to sputtering by ions within Jupiter's plasma torus. This material loss, composed of ionized and atomic sulfur, oxygen, and chlorine, along with atomic sodium and potassium, forms Io's extended neutral clouds and corona. Particles in the plasma torus, a doughnut-shaped structure of ionized atoms, co-rotate with Jupiter's magnetosphere, facilitating the removal of neutral atoms and molecules from Io's atmosphere. Variations in magnetospheric plasma, linked to Io's volcanic activity, have been observed far into Jupiter's magnetotail.

Jupiter's magnetic field interacts strongly with Io, producing distinctive auroras primarily near Io’s equator. The Io flux tube generates an electric current that results in auroral glows at Jupiter's polar regions and in Io's atmosphere. These auroras fluctuate with changes in Jupiter’s magnetic field and are observable in various colors due to different emitting gases: molecular sulfur dioxide causes blue glows, while atomic oxygen and sodium contribute to red glows. Eclipses cause noticeable variations in auroral brightness, reflecting changes in atmospheric gas density and electron interactions. These auroral emissions illuminate not only Io's atmospheric composition but also its dynamic interaction with Jupiter’s magnetosphere.

**3. Titan**

Although similar in size to Ganymede and Calisto, Titan, Saturn's largest moon, appears to be a completely different world. Titan is the only moon in the Solar System with a substantial atmosphere. The presence of an atmosphere on Titan was first suggested by limb darkening observations in 1908 and later confirmed by the discovery of $CH_4$ in 1944. The flyby of Pioneer 11 in 1979 discovered abundant atmospheric aerosols. The Voyager flybys in the early 1980s confirmed Titan's $N_2$-dominated atmosphere. The Cassini spacecraft carried the Huygens probe from 2005 to 2017 for detailed Titan characterization with remote sensing and in situ analysis.

A schematic view of the important processes in Titan’s atmosphere is shown in Figure 3. The atmosphere is primarily composed of $N_2$, a few percent of $CH_4$, and about 0.1% of hydrogen ($H_2$). Titan's atmosphere is enriched with a variety of hydrocarbons and nitriles, following $CH_4$ photolysis in the thermosphere, catalytic reactions in the stratosphere, and $N_2$ dissociation caused by UV photons and energetic electrons. To date, Titan's atmosphere has been identified to contain 10 other hydrocarbons ($C_2H_2$, $C_2H_4$, $C_2H_6$, c-$C_3H_2$, $CH_2CCH_2$, $CH_3CCH$, $C_3H_6$, $C_3H_8$, $C_4H_2$, c-$C_6H_6$) and 8 cyanides (HCN, HNC, $HC_3N$, $C_2N_2$, $CH_3CN$, $C_2H_3CN$, $C_2H_5CN$, $CH_3C_3N$). Some of their

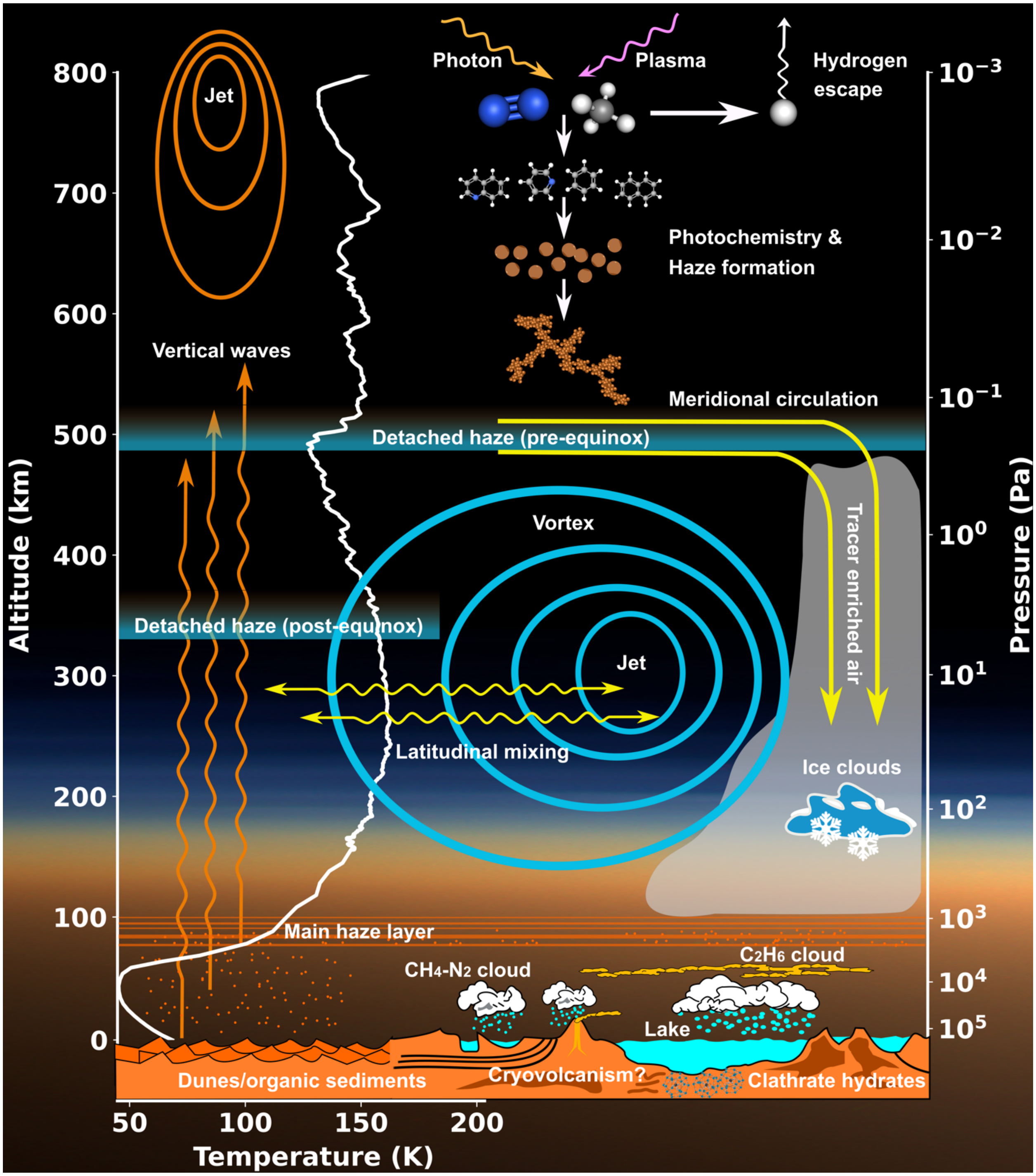

***Figure 3.*** *Schematic view of Titan's atmospheric processes. In the upper atmosphere, $N_2$ and $CH_4$ undergo photochemistry and ion-chemistry, leading to the formation of hydrocarbons, nitriles, and aerosols. Methane and hydrogen escape into space. Vertically propagating waves induce significant temperature oscillations and might cause the intense thermospheric jet. The middle atmosphere is characterized by a superrotating zonal wind. Both temperature and chemical tracers display strong latitudinal and seasonal variations. Tracer distributions are influenced by photochemistry, meridional circulation, polar vortex, and latitudinal mixing.*

*After equinox, the detached haze layer descends from 500 km to about 350 km. Ice clouds form in the descending branch of the winter polar stratosphere. In the lower atmosphere, organic haze particles settle gravitationally from the main haze layer, forming equatorial dunes. The methane-based hydrological cycle shows large convective clouds, episodic storms, and intense precipitation. $CH_4$ is replenished through the evaporation of polar lakes and seas, as well as potential subsurface sources like clathrate hydrates and cryovolcanism. The temperature profile was from the Huygens probe (Fulchignoni et al. 2005). This figure was inspired from prior works by Teanby et al. (2008), Hörst (2017), Hayes et al. (2018), and Nixon (2024). The background image source: the Cassini-Huygens mission (PIA06236).*

isotopologues (e.g., $^{13}C$, $^{15}N$), have also been detected. Ethane ($C_2H_6$) is the most abundant hydrocarbon other than $CH_4$ and hydrogen cyanide (HCN) is the predominant nitrile. Noble gases like argon ($^{36}Ar$, $^{40}Ar$) and neon ($^{22}Ne$) have been measured. A few simple oxygen-bearing species including CO, $CO_2$, and $H_2O$ have also been detected, but many expected chemical families of nitrogen- and oxygen-bearing molecules remain elusive. Other than the CHNO-bearing species, only the upper limits of $PH_3$ and $H_2S$ were obtained from Cassini observations. In addition, Titan has a substantial ionosphere above 1000 km, with large abundant ions with masses exceeding 300 Daltons, and some negative ions with masses that exceed 10,000 Daltons.

Titan is the only known extraterrestrial body with liquid on its surface. With a surface pressure of 1.5 bar and a temperature of 94 K, Titan's conditions are close to the triple point of methane, enabling a hydrological methane cycle in the lower atmosphere similar to Earth's water cycle. The surface of Titan is warmed by a greenhouse effect caused by $CH_4$ and collision-induced absorption ($N_2$-$N_2$, $N_2$-$CH_4$, $N_2$-$H_2$) in the atmosphere. The anti-greenhouse effect occurs as the upper atmospheric haze absorbs visible and UV radiation but remains more transparent at infrared wavelengths, allowing the surface to emit and cool into space. Titan is unique due to the existence of liquid on Titan's surface and the complex interactions between the atmosphere and surface. The surface terrain, predominantly sculpted by erosive processes involving liquids and winds, features dunes at low latitudes, hydrocarbon seas and lakes at high latitudes, and a river network that resembles Earth's hydrological systems.

Similar to Earth, the vertical temperature structure in Titan's atmosphere can be characterized by several layers from bottom to top: the troposphere, stratosphere, mesosphere, and thermosphere (Figure 3). The temperature decreases with altitude in the troposphere, which appears to be statically stable, with a tropopause temperature of about 70 K at an equatorial altitude of 44 km. The lower stratosphere exhibits high static stability with a temperature gradient of about 1 K per km. The radiative energy transfer in the stratosphere is governed by the absorption of solar radiation by haze in visible and UV wavelengths, and by methane in the near-IR bands, along with cooling by hydrocarbons such as $C_2H_2$ and $C_2H_6$. The stratopause of Titan ranges from 260 km with a temperature of 187 K, according to the Huygens Atmospheric Structure Instrument (HASI), to around 312 km with a temperature of 183 K, based on Cassini remote sensing.

Above the stratopause, the temperature decreases with altitude in the mesosphere, where solar absorption by thin haze becomes less important than that by the $CH_4$ infrared bands. Non-LTE effects begin to be significant for hydrocarbon cooling above 400 km. Titan's mesopause temperature reaches about 140-150 K at 500-800 km, above which the temperature rises in the thermosphere, where significant UV heating by methane is offset by efficient cooling from far-infrared HCN rotational lines, with thermal conduction playing a minor role. From HASI observations, Titan's upper atmosphere exhibits significant oscillations, likely due to wave propagation. The data suggested the presence of an approximately isothermal region around 170 K from 500 to 1100 km, superimposed with 10 K thermal waves. At altitudes above 1000 km, the temperature profile may not be isothermal due to adiabatic cooling potentially influencing the temperature distribution and a possible slow hydrodynamic escape from Titan's atmosphere. The exobase is located at around 1500-1600 km, where gases can freely escape to space.

The chemical species in Titan's atmosphere exhibit significant vertical variation, primarily influenced by vertical transport and the balance of chemical sources and sinks. The dominant species, $N_2$, is roughly well-mixed throughout the atmosphere. However, the mole fraction of $CH_4$ starts at about 5% near the surface and decreases to 0.014 in the stratosphere due to cold trapping via cloud formation in the troposphere. Above the homopause (around 800 km), $CH_4$ separates from nitrogen and its mole fraction increases to 12% at the exobase. $N_2$ and $CH_4$ are transported upwards to be dissociated, initiating the photochemistry in Titan's atmosphere.

The photochemical products of $N_2$ and $CH_4$, including hydrocarbons and nitriles, peak in their production region in the upper atmosphere and are transported downward. In general, the gas-phase chemical processes on Titan involve the irreversible transformation of $CH_4$ and $N_2$ into a variety of complex hydrocarbons and nitriles, which contribute to the formation of aerosol particles and condensational clouds in the lower atmosphere. These condensed phases eventually integrate into the hydrological system or contribute to the formation of equatorial dunes on Titan's surface. The photochemical reactions also lead to extensive production of atomic and molecular hydrogen, which can ascend and readily escape from the atmosphere. The escape of $H_2$ is limited by the maximum diffusion rate of $H_2$ and the average escape rate is about $9.2 \times 10^{27}$ $H_2$ $s^{-1}$, while the H escape rate might be one order of magnitude smaller. The observed steep gradient of $CH_4$ in the upper atmosphere suggested a large escape rate of $2 \times 10^{27}$ $CH_4$ $s^{-1}$, but it could also be due to an unknown chemical loss instead of escape loss.

A comprehensive understanding of the organic chemistry in Titan's atmosphere remains elusive. While the neutral chemistry on Titan is relatively well studied, the ion chemistry introduces significant uncertainties. In Titan's atmosphere, the neutral and ion chemistries are intimately linked. The state-of-the-art coupled ion-neutral photochemical model includes about 80 neutral species and 200 ions up to m/z ~100, with nearly 2000 ion-neutral reactions. A summary of the major sources and sinks of important species on Titan is provided in Table 2.

The dissociation of $N_2$ and $CH_4$ primarily occurs above 700 km, facilitated by EUV and UV solar photons on the dayside and electrons from the Saturnian magnetosphere on the nightside. Dissociation may also take place in the lower atmosphere due to energetic protons, ions, galactic cosmic rays, and through catalytic processes. $CH_4$ dissociation yields radicals such as CH, $CH_2$, and $CH_3$. The self-recombination of $CH_3$ radicals produces $C_2H_6$, the main photochemical species. $C_2H_4$ is generated from the reaction of methane with its derived radicals (CH + $CH_4$) and through radical-radical reactions (3$CH_2$ + $CH_3$). The subsequent photodissociation of $C_2H_4$ is the primary source of $C_2H_2$, with an additional contribution from hydrogen abstraction from $C_2H_3$ in the upper atmosphere. The photolysis of $C_2H_2$ produces the $C_2H$ radical, which is a key precursor for producing $C_4H_2$ ($C_2H$ + $C_2H_2$) and $C_4H_4$ ($C_2H$ + $C_2H_4$). In the stratosphere, the radicals $C_2H$ and $C_2$ can also dissociate $CH_4$ and recycle $C_2H_2$ back. $C_3$ compounds, such as $C_3H_8$ and $CH_3CCH$, are synthesized via the insertion of a $C_1$ radical into a $C_2$ molecule.

***Table 2.*** *Important Chemical Production and Loss Pathways for Abundant Neutral Species in the Atmosphere of Titan, based on Hörst (2017), Vuitton et al. (2019), and Nixon (2024).*

| Molecule | Production | Loss |
|---|---|---|
| **Nitrogen ($N_2$)** | Originated from $NH_3$ | $N_2$ + hv → $N(^2D)/N(^4S)/N^+$ |
| **Methane ($CH_4$)** | Clathrates/hydrothermal reactions | $CH_4$ + hv → $CH_3$ + H<br>Escape |
| **Hydrogen ($H_2$)** | $CH_4$ photolysis | Escape |
| **Carbon Monoxide (CO)** | O + $CH_3$ → $H_2CO$ + H<br>$H_2CO$ + hv → CO + $H_2$/2H | CO + OH → $CO_2$ + H |
| **Ethane ($C_2H_6$)** | $CH_3$ + $CH_3$ + M → $C_2H_6$ + M | Condensation |
| **Acetylene ($C_2H_2$)** | $C_2H$ + $CH_4$ → $C_2H_2$ + $H_2$ | $C_2H_2$ + hv → $C_2H$ + H<br>Condensation |
| **Propane ($C_3H_8$)** | $CH_3$ + $C_2H_5$ + M → $C_3H_8$ + M | Condensation |
| **Ethylene ($C_2H_4$)** | CH + $CH_4$ → $C_2H_4$ + H<br>$^3CH_2$ + $CH_3$ → $C_2H_4$ + H | $C_2H_4$ + hv → $C_2H_2$ + $H_2$/2H |
| **Hydrogen Cyanide (HCN)** | $N(^4S)$ + $CH_3$ → $H_2CN$ + H<br>$H_2CN$ + H → HCN + $H_2$ | Condensation |
| **Carbon Dioxide ($CO_2$)** | CO + OH → $CO_2$ + H | Condensation |
| **Methylacetylene ($CH_3CCH$)** | CH + $C_2H_4$ → $CH_3CCH$ + H | $CH_3CCH$ + hv → $C_3H_3$ + H<br>H + $CH_3CCH$ → $C_3H_5$ |
| **Diacetylene ($C_4H_2$)** | $C_2H$ + $C_2H_2$ → $C_4H_2$ + H | $C_4H_2$ + hv → $C_4H$ + H |
| **Propene ($C_3H_6$)** | H + $C_3H_5$ → $C_3H_6$<br>$CH_3$ + $C_2H_3$ + M → $C_3H_6$ + M | H + $C_3H_6$ → $C_3H_7$ |
| **Benzene ($C_6H_6$)** | $C_6H_7^+$ + e → $C_6H_6$ + H | $C_6H_6$ + hv → $C_6H_5$ + H<br>Condensation |

The dissociation of $N_2$ yields products like N(4S), N(2D), $N^+$, and possibly $N_2^+$. An important species, $H_2CN$, is produced in two ways: firstly, through the reaction of $N(^4S)$ with the $CH_3$ radical, and secondly, $N(^2D)$ + $CH_4$ yields $CH_2NH$, which reacts with H to produce $H_2CN$. $H_2CN$ then either combines with H or undergoes direct photodissociation to form HCN, the predominant nitrile. $N(^4S)$ can also react with $^3CH_2$ radicals to form HNC. The photolysis of HCN generates CN radicals, which are highly reactive and can react with HNC to form $C_2N_2$, with $C_2H_2$ to produce $HC_3N$, and with $C_2H_4$ to form $C_2H_3CN$.

In Titan's upper atmosphere, ion chemistry predominates. $N^+$, $N_2^+$, and $CH_4^+$ are the primary ionization products from $N_2$ and $CH_4$, giving rise to various secondary ions. The principal pathway for the loss of positive ions is through radiative or dissociative electron recombination. Negative ions, unexpected in early models, appeared to be abundant in Titan's atmosphere. These negative ions are likely generated through mechanisms such as photoionization, dissociative electron attachment, and radiative electron attachment. Their loss processes include photodetachment, associative detachment with neutrals, and recombination with positive ions. Electron recombination reactions play a crucial role in synthesizing some neutral species, such as benzene ($C_6H_6$) and potentially larger polycyclic aromatic hydrocarbons (PAHs) in the upper atmosphere, where the low pressure limits three-body collisions. The observed abundance of $C_6H_6$ is attributed to electron recombination with the $C_6H_7^+$ ion.

The measured abundances of oxygen-bearing species (CO, $CO_2$, $H_2O$, and precipitating $O^+$ ions) in Titan's atmosphere have not been well explained. To supply the necessary OH radicals that react with CO to form $CO_2$, an external influx of oxygen into the upper atmosphere seems necessary. This oxygen source is thought to come from micrometeorite ablation or the influx from the plumes of Saturn's moon Enceladus, which is also believed to contribute to Saturn's E ring. The chemical lifetimes of CO (~0.1 billion years), $CO_2$ (~0.1 million years), and $H_2O$ (~100 years) differ significantly, suggesting that variable influxes of external oxygen reach Titan on different timescales.

Titan's atmosphere, composed of $N_2$, $CH_4$, and CO, serves as an ideal factory for the photochemical production of aerosol (or haze) particles, which are ubiquitous below 300 km (Figure 3). Cassini observations have unveiled several aerosol layers extending from the surface up to approximately 1000 km. The Descent Imager/Spectral Radiometer (DISR) on the Huygens probe characterized the primary haze layer below roughly 150 km, identifying several thin layers and one notably thick layer. Limb emission and solar occultation observations from Cassini have probed the haze extinction profile between 100-500 km in the infrared spectrum, while stellar occultation data have resolved the 400-1000 km region in the far ultraviolet (FUV) with a vertical resolution of 10-50 km. The UV extinction profile shows two local maxima: one below 500 km and another around 700-800 km. The locations of the "detached haze layers" at around 500-600 km exhibit significant variations in latitude and season.

The chemical nature of the aerosols is not well constrained. Beyond benzene, Titan's chemistry yields complex macromolecules and heavy ions above 1000 km, which are thought to be precursors to aerosols, including aliphatic copolymers, nitriles, aromatics, and polyynes. Laboratory experiments have successfully produced haze analogues similar to those in Titan's atmosphere, often referred to as tholins. However, these laboratory analogues lack the strong C-H stretching feature at 3.4 µm observed in Cassini spectra. Instead, they exhibit N-H and C-N features at 3 and 4.6 µm, which are absent in the observed spectra, suggesting that Titan's aerosols might be less nitrogen-rich than their laboratory counterparts. Pyrolysis experiments on the Huygens probe might have detected $NH_3$ and HCN in the aerosol composition, but the interpretation remains ambiguous. The aerosol particles are not compact spheres; Cassini imagery indicates that the aerosols are fractal in nature, with a fractal dimension of about two. Once the precursors form in the upper atmosphere, the particles begin to nucleate into tens-of-nanometers-sized monomers at roughly 800 km. These monomers then settle and grow by coagulation to form fractal aggregates, consisting of approximately 4000 monomers each, about 40-50 nm in size, within the main aerosol layer.

Aerosols largely influence Titan's atmosphere. They shield against UV radiation, affecting photochemistry, and play a crucial role in heterogeneous reactions, such as recombining H atoms on their surface to aid in the synthesis of complex hydrocarbons. Aerosols also shape the radiative flux within Titan's atmosphere through their "anti-greenhouse" effect and can locally heat the atmosphere close to the magnitude of $CH_4$ heating in the troposphere and stratosphere. Aerosols could also act as condensation nuclei for organic clouds observed on Titan, including convective $CH_4$ clouds in the troposphere, stratiform $C_2H_6$ clouds between 30-65 km, and cirrus-like ice clouds in the stratosphere such as HCN, $HC_3N$, and $C_4N_2$ (Figure 3). The aerosols and ices that settle on the ground may undergo further surface processing. Through hydrolysis, laboratory-synthesized “tholins” can produce prebiotic molecules, including amino acids, which holds considerable interest in the field of astrobiology.

The temperature and composition of Titan's atmosphere exhibit variability both spatially and temporally, influenced by variations in solar insolation, atmospheric chemistry, large-scale circulation, atmospheric waves, and Titan's position within Saturn's magnetosphere (Figures 3 and 4). The solar insolation varies significantly with the seasons due to Titan's obliquity of 27 degrees. With a slow rotation period of 16 days, Titan's climate resembles a tropical climate. The mean meridional circulation is dominated by a pole-to-pole Hadley-type cell with ascending motions in the summer hemisphere and subsiding motions in the winter polar region. Following the equinoxes, the circulation reverses, forming two equator-to-pole cells that persist for several Earth years. This meridional circulation redistributes heat across latitudes in the lower atmosphere, where the atmospheric radiative timescale exceeds the length of a Titan year (29.5 Earth years). Surface temperatures are almost uniform up to about 35 degrees latitude and gradually decrease by about 3 K from the equator to the winter pole and by 1 K to the summer

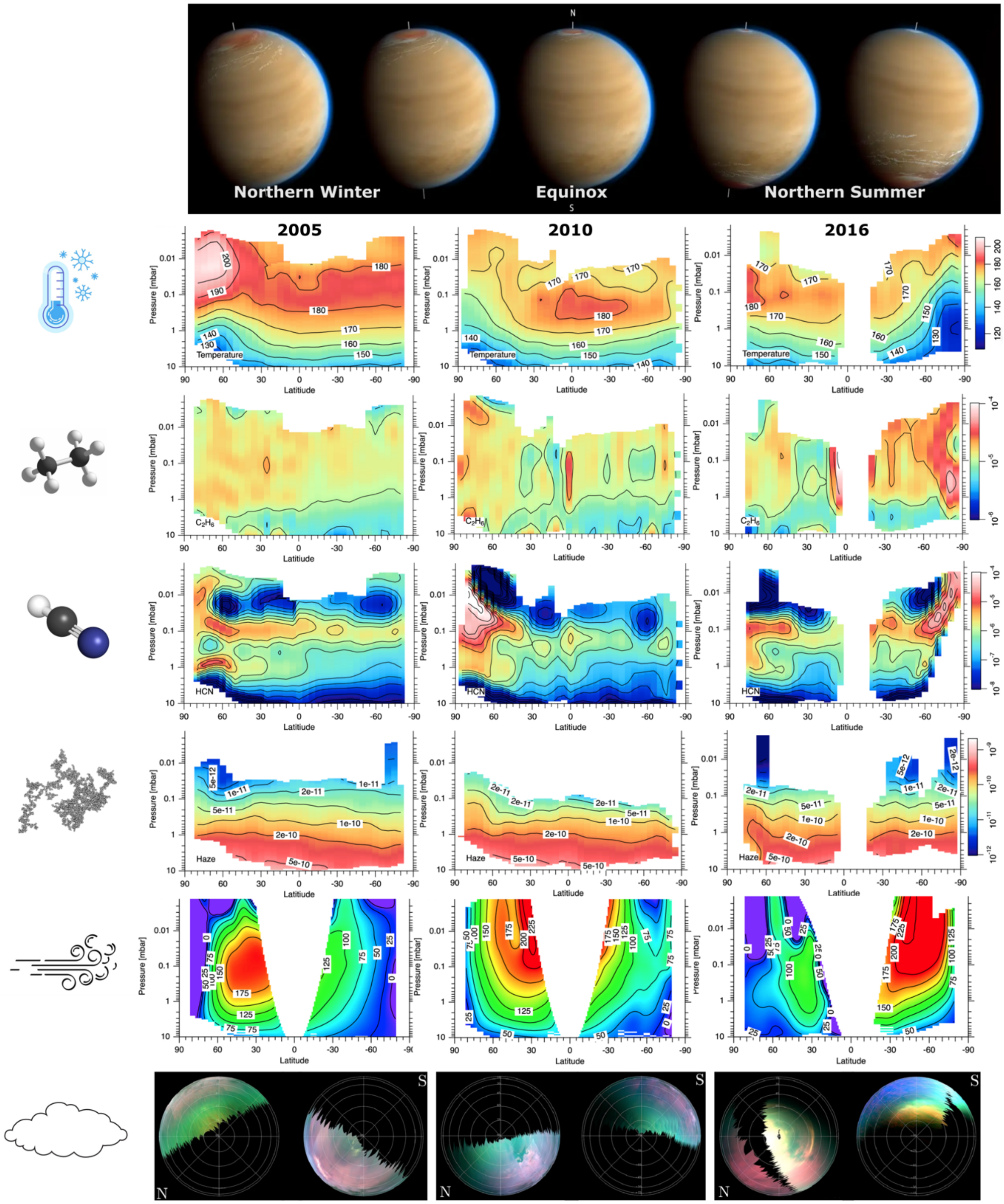

***Figure 4.*** *The compilation of Cassini observations on the seasonal variations in Titan's atmosphere. The top row is an artistic depiction of Titan's seasonal change from northern winter to northern summer, as indicated by the axis tilt (adapted from NASA image PIA16481). The images show typical features such as the polar hood (white streaks), the polar stratospheric "hot spot" (red), and the detached haze layer (blue). Based on Cassini observations, the polar hood and hot spot appear in the northern polar region when it is*

*pointed away from the sun. At equinox, both hemispheres receive equal sunlight, leading to spring in the north and autum. nAt equinox, both hemispheres receive equal sunlight, leading to spring in the north and autumn in the south. Post-equinox, after 2011, trace gases built up over the north pole despite the dissipation of the vortex and hot spot, with similar features emerging at the south pole. These observations suggest a large-scale reversal in Titan's atmospheric circulation, with upwelling in the summer hemisphere and downwelling in the winter hemisphere. Rows two through six illustrate retrieved latitudinal distributions of temperature, $C_2H_6$, HCN, haze, and the thermal wind, respectively (figures adapted from Achterberg, R.K., 2023. Temporal evolution of Titan's stratospheric temperatures and trace gases from a two-dimensional retrieval of Cassini composite infrared spectrometer data 4, 140.). The corresponding altitudes of the pressure levels are shown in Figure 3. Columns from left to right represent Cassini observations from 2005 (northern winter), 2010 (equinox), and 2016 (northern summer), respectively. The last row illustrates the change in cloud patterns at the northern ("N") and southern ("S") poles (figures adapted from Le Mouélic, S., Rodriguez, S., Robidel, R., Rousseau, B., Seignovert, B., Sotin, C., Barnes, J.W., Brown, R.H., Baines, K.H., Buratti, B.J., Clark, R.N., Nicholson, P.D., Rannou, P., Cornet, T., 2018. Mapping polar atmospheric features on Titan with VIMS: From the dissipation of the northern cloud to the onset of a southern polar vortex 311, 371–383.). These data collectively reveal that in the winter hemisphere, tracers and clouds are more enriched, and the stratospheric jet is more intense. During the equinox, distributions tend to be more symmetric.*

pole. The near-surface two planetary boundary layers—a seasonal one at 2 km and a diurnal one peaking at 800 m—indicates variations in both daily and annual cycles.

The Hadley cell may extend up to 600 km (around 0.05 Pa) above the surface. The radiative response time is shorter in the stratosphere, and atmospheric variability is much stronger. Due to a strong mid-latitude jet and a large temperature gradient near the winter pole, Titan exhibits an intense winter polar vortex (Figure 3). In the vortex, strong subsidence combined with concentrated aerosol heating results in high temperatures in the upper stratosphere (around 1 Pa) exceeding 200 K, marking the warmest areas in Titan's atmosphere. This phenomenon has been observed at both the northern and southern poles of Titan in different seasons. Many hydrocarbons and nitriles, as well as the polar clouds, also exhibit higher concentrations at the winter pole than at lower latitudes or in the summer hemisphere (Figure 4). This pattern can be explained by the strong subsidence of species from their source regions in higher altitudes to the winter polar night, where the photochemical destruction is weak. Shorter-lived species (e.g., $C_2H_4$, $C_3H_4$, $C_4H_2$, $C_6H_6$, $HC_3N$) exhibit stronger variations compared to longer-lived species (e.g., $C_2H_2$, $C_2H_6$, $C_3H_8$, $CO_2$).

In addition to tracer gases, the aerosols on Titan exhibit significant variability. The winter hemisphere appears hazier than the summer hemisphere. The detached haze layer consists of two types of particles: a group of 50 nm particles and another group of larger particles. The larger particles are likely transported upward by large-scale circulation from the main haze layer. During the equinox, when the circulation switches directions, the vertical wind changes. Consequently, the detached layer was observed to drop from 500 km to about 350 km (Figure 3). Above the detached haze layer, some layered structures are observed in the hazes, which might be caused by gravity waves, similar to those occurring in Pluto's atmospheric hazes.

The zonal wind speed on Titan varies significantly with altitude. At the surface level, the lack of wave activities detected in Titan's lakes implies a weak surface wind (less than 1 m $s^{-1}$). The orientation and shape of Titan's dunes suggest either a reversal from westward to eastward winds or the generation of strong eastward gusts by equinoctial storms. In Titan's lower troposphere, the circulation is in the geostrophic regime. The zonal wind increases to approximately 40 m $s^{-1}$ near 60 km and then decreases to about 5 m $s^{-1}$ at around 75 km. The wind speed increases again with altitude, reaching approximately 200 m $s^{-1}$ at about 200 km in the winter hemisphere. With strong superrotating winds, the stratosphere maintains cyclostrophic balance, similar to the superrotating region of Venus. The wind speed appears to decrease to 60 m $s^{-1}$ near 450 km, exhibiting significant latitudinal and seasonal variations. Strong equatorial superrotating winds then increase with altitude in the thermosphere above 600 km, reaching about 340 m $s^{-1}$ at 1000 km. Given Titan's cold temperature, this wind speed is considered supersonic (Figure 3).

General circulation models (GCMs) suggest that the origin of the superrotation in the middle and upper atmosphere of Titan might be attributed to the Gierasch-Rossow-Williams mechanism, which was proposed to explain the superrotating wind on Venus. In Titan's atmosphere, the superrotation can be sustained by the upward transport of angular momentum via mean meridional circulation, complemented by equatorially transported momentum from barotropic waves propagating in the winter stratosphere. Equatorial Kelvin-like waves and Rossby-like waves play crucial roles in wind generation. Moreover, the distribution of radiatively active tracers in the stratosphere, along with seasonal variations, greatly influences the superrotation. The mechanism for the thermospheric jet remains unclear. It has been suggested that the jet could be driven either by upward-propagated waves from the upper stratosphere and mesosphere or through magnetospheric-ionospheric interactions.

Titan's atmosphere might also exert torques on the entire planetary body. It has been suggested that angular momentum transfer between Titan's dense atmosphere and its surface could cause non-synchronous rotation of Titan with Saturn, which may be further influenced by the seasonal shift of the Hadley circulation. While Cassini observations did not find evidence for non-synchronous rotation, they did reveal that the axis of symmetry for stratospheric temperature and winds, including the polar vortex, is tilted several degrees away from the north pole. This tilt appears to be fixed in the inertial frame without seasonal variation, and the underlying mechanism remains elusive.

The clouds on Titan are composed of methane, ethane, and other condensable simple organics, but the cloud distribution is complex (Figure 3). Convective methane clouds typically occur in the summer hemisphere around the pole and mid-latitudes, while condensational ethane clouds and nitrile clouds predominantly form around the winter pole. During equinoxes, clouds have been observed at both poles. This pattern might be largely driven by surface warming, general circulation, and atmospheric waves. Large cloud systems, covering up to approximately 10% of the disk, occur every 3 to 18 months on Titan, increasing the brightness of Titan's disk by several tens of percent. These storms could trigger Rossby waves and form large-scale clouds in other locations. Wave-induced clouds are particularly important in the formation of tropical clouds near the equinox. For example, low-latitude storms have been linked to equatorial Kelvin waves that organize convection into chevron-shaped storms. However, determining whether clouds are primarily triggered by wave activity or by deep convection remains a challenge. To date, lightning has not been detected on Titan.

The methane cycle on Titan is analogous to Earth's hydrological cycle, involving methane evaporation from the surface, cloud formation, and precipitation back to the surface. Titan's surface is predominantly dry, with lakes and seas constituting about 1% of its total surface area. Notably, 97% of the surface liquids are located in the north polar region, with 80% concentrated in three large seas. Observations suggest that several meters of liquid evaporate from the polar lakes during the summer. Intermittent rainstorms, typically lasting a few hours, can deposit about a few meters of rainfall per storm in the polar regions. On average, it was estimated that there are 10-100 storms per season in the summer polar region, one storm per season in the equatorial region around the equinox, and 1-10 storms per season at the summer middle latitudes. The estimated average methane precipitation rates generally range from 0.001 to 0.5 cm per Earth year.

Methane is transported across latitudes by large-scale circulation, which is crucial for explaining the observed mid-latitude cloud systems. The current orbital configuration of the Saturn-Titan system, with the southern summer solstice coinciding with perihelion, results in a net northward transport of methane due to the reversal of pole-to-pole Hadley circulation, estimated at 1000 $km^3$ per Titan year. As orbital parameters change, such as Saturn's perihelion precession in a 45-kyr cycle, this net transport direction may oscillate over 10–100 kyr periods, analogous to Earth's Croll–Milankovitch cycles. Given that the total volume of methane in Titan's surface liquids is about 70,000 $km^3$, the methane flux could significantly alter the polar basins and lakes over Milankovitch timescales, potentially explaining the hemispheric asymmetry in Titan's surface liquid distribution. In addition, atmospheric circulation models coupled with surface hydrology also suggest that topography, as well as surface and subsurface hydrology, could influence liquid distribution.

Unlike Earth, where most water is condensed in the oceans, Titan's atmosphere contains seven times more methane than its surface liquids. Despite cloud formation, methane is not globally cold trapped in the troposphere but is transported upward and irreversibly destroyed by photochemistry in the upper atmosphere. The lifetime of methane in Titan's atmosphere is about 30 Myr, suggesting that the current atmospheric methane is relatively recent. This is supported by the $^{12}C/^{13}C$ isotope ratio observations, which show no large fractionation due to atmospheric loss. It is hypothesized that methane might be replenished by sources other than the surface liquids, potentially trapped as clathrate hydrates within Titan's interior and released through ammonia-water cryovolcanism or substituted by ethane. The detection of $^{40}Ar$, a decay product of $^{40}K$, indicates the presence of outgassing in Titan's history.

The origin of Titan's atmosphere remains elusive. The oxygen in CO on Titan might have an external source, indicating that it was not necessary to accrete abundant CO during Titan's formation. The lower-than-solar $^{36}Ar/N$ ratio strongly suggests that Titan's nitrogen originated as $NH_3$. The $^{14}N/^{15}N$ ratio on Titan is much lower than that on Saturn, suggesting that the original $NH_3$ was acquired from the protosolar nebula, not the subsaturnian nebula. $NH_3$ may have been converted to $N_2$ within Titan's interior, through impact-induced chemistry, or via photolysis in the atmosphere. The origin of methane-trapped clathrates remains unknown. If the clathrates are primordial—accreted during Titan's formation—then Titan should have Krypton and Xenon levels much higher than the detected upper limits. Conversely, if methane is produced internally from low-temperature hydrothermal reactions, the D/H ratio in equilibrium with water inside Titan would be significantly lower than that in the plume of Saturn's moon Enceladus, another moon of Saturn.

As an organic chemical factory, Titan's atmosphere offers valuable insights into conditions on early Earth before atmospheric oxygen levels increased. For example, an analogous organic haze in Earth's ancient atmosphere may have protected primordial life by shielding it from UV radiation from the early sun. Titan's climate also demonstrates important aspects of atmospheres in a moist greenhouse state, where moisture is not cold trapped in the troposphere but can reach the stratosphere and potentially escape into space. The moist or runaway greenhouse scenario is relevant to Earth's future climate under a brightening sun, as well as to Venus's past climate before it lost its water. As the only dense terrestrial atmosphere in the outer Solar System, Titan offers unique intercomparison opportunities to understand terrestrial climates within the inner Solar System.

## 4. Pluto

The presence of an atmosphere on Pluto was first postulated when $CH_4$ ice was detected on its surface in 1976, with confirmation following a stellar occultation in 1988. Subsequent detections of $N_2$ and CO ices refined our understanding of the atmosphere's composition, revealing it to be predominantly $N_2$, with minor constituents of $CH_4$ and CO. The 2015 flyby by NASA's New Horizons mission provided the most detailed information to date about Pluto's atmosphere, though it represented only a snapshot, as the atmosphere is expected to change significantly over time.

During the 2015 encounter, New Horizons observed that Pluto's atmosphere was composed of about 99% $N_2$, 0.5% $CH_4$, and 0.05% CO. Additionally, several hydrocarbons, including $C_2H_2$, $C_2H_4$, and $C_2H_6$—photochemical products of $CH_4$—were identified. HCN and further traces of CO were detected through near-infrared reflection spectra and radio-telescope observations. It was also found that Pluto's atmosphere is characterized by numerous fine haze layers, which are crucial for energy transport and chemical processes within the atmosphere. Unlike Pluto, its largest moon, Charon—which is about half the size of Pluto—shows no signs of an atmosphere. However, the north pole of Charon exhibits a red coloration, possibly indicating a transfer of hydrocarbon gases from Pluto that were subsequently altered by UV radiation, resulting in a reddish deposit on Charon's polar surface.

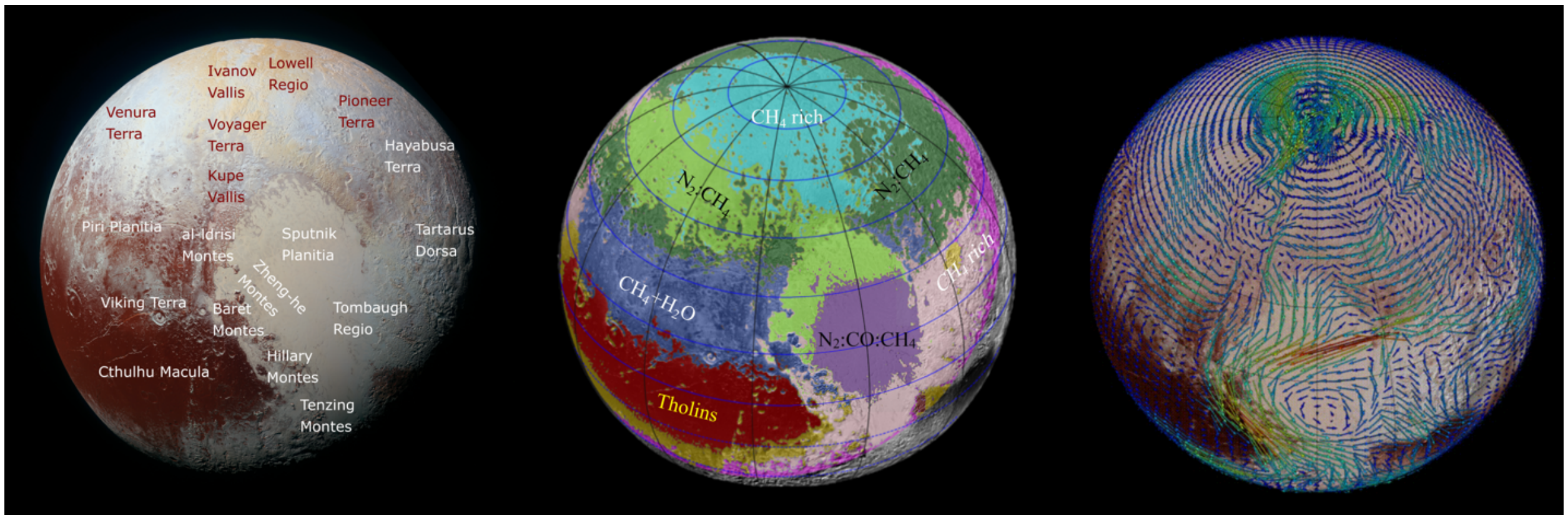

***Figure 5.*** *Global distributions of geological units (left), surface ices (middle), and wind patterns (right) on Pluto. The color map of Pluto is from the New Horizons mission (NASA image: PIA19708). The surface ice composition map was adapted from Emran, A., Dalle Ore, C.M., Ahrens, C.J., Khan, M.K.H., Chevrier, V.F., Cruikshank, D.P., 2023. Pluto's surface mapping using unsupervised learning from near-infrared observations of LEISA/Ralph. The Planetary Science Journal, 4, 15. The global wind map was simulated from a Pluto GCM (Bertrand, T., Forget, F., Schmitt, B., White, O.L. and Grundy, W.M., 2020. Equatorial mountains on Pluto are covered by methane frosts resulting from a unique atmospheric process. Nature communications, 11(1), p.5056.); used by permission of simulation author Tanguy Bertrand.*

Although Pluto's atmospheric composition resembles that of Titan, Pluto's atmosphere is much thinner and colder, with a surface pressure on the order of 0.1-1 Pa and a temperature ranging from 38-55 K. Pluto's semi-major axis is approximately 39.5 AU, resulting in an incoming solar flux less than one-tenth of that reaches Titan. This places Pluto's atmosphere in a "snowball" version of Titan's, characterized by a significantly collapsed atmosphere. On Pluto, the atmospheric composition and pressure are influenced by solar radiation and surface ices. The entire atmosphere is part of a comprehensive volatile cycle that is controlled by the sublimation and condensation processes of the surface ices composed of $N_2$, $CH_4$, and CO.

Data from the New Horizons flyby in 2015 revealed that the distribution of surface ices on Pluto is strongly correlated with geological features and latitude (Figure 5). Generally, $N_2$ ice is found primarily in topographical lows, while $CH_4$ ice tends to concentrate at higher altitudes. This altitude segregation, although not fully understood, suggests that $N_2$-rich ice tends to accumulate and be more thermodynamically stable in warmer, lower-altitude, high-pressure regions. Conversely, $CH_4$-rich ice may accumulate at higher altitudes due to atmospheric circulation that transports it upward, followed by condensation at mountain summits. The largest reservoir of $N_2$ ice is located in Sputnik Planitia, the western lobe of the Tombaugh Regio. This extensive basin, centered at 15°N and 180°E, measures approximately 1200 km by 1400 km and is filled several kilometers deep with primarily $N_2$-rich ice. $CH_4$-rich ice and CO ice are also present in Sputnik Planitia.

Adjacent to this area in the equatorial zone is the Cthulhu Macula, which features dark, volatile-free areas with reddish surface materials. However, $CH_4$-rich ice is detected on high-altitude ridges and north-facing slopes, as well as in the lower to middle northern latitudes (10°N–35°N). The mid-latitudes (35°N–55°N) again exhibit a predominance of $N_2$-rich ice at low altitudes, with trace amounts of CO. Beyond 60°N, $CH_4$-rich ice dominates the northern high latitudes. The distribution of Pluto's ice exhibits strong seasonal behavior, as detailed later.

To a first approximation, the main atmospheric components, $N_2$ and $CH_4$, should be in vapor-pressure equilibrium with their respective ices. If the $CH_4$ and $N_2$ ices were uniformly distributed over the surface, thermodynamics predicts that the $CH_4$ mixing ratio over the surface should be less than 0.01% at temperatures around 36-40 K, which is much lower than observed. This discrepancy could be resolved if the $N_2$ and $CH_4$ ices are not uniformly mixed globally, which seems likely given the discovery of almost pure $CH_4$ ices in the north polar region and other areas on Pluto. The $CH_4$ sublimated from these warmer, pure $CH_4$ patches could explain the high mixing ratio of $CH_4$ in Pluto's atmosphere, underscoring the importance of understanding the heterogeneous distribution of ices and the volatile cycle on Pluto.

Radio occultation observations during the New Horizons flyby determined Pluto's mean radius (1188.3 ± 1.6 km) and vertical temperature profiles derived from the air density distribution (Figure 6). Two temperature profiles, from the surface up to 100 km, were obtained during the ingress and egress phases, respectively. The ingress occultation, occurring above southern Sputnik Planitia, yielded surface temperature and pressure consistent with vapor-pressure equilibrium over $N_2$ ice (1.28 ± 0.07 Pa and 38.9 ± 2.1 K). A distinctive boundary layer, characterized by a negative temperature gradient with altitude, was evident in the lower 4 km over the $N_2$ ice surface. This boundary layer is likely caused by the $N_2$ mass flux sublimated from the ice during the day, but whether the lapse rate follows a dry or moist $N_2$ adiabat remains unclear due to limited vertical resolution. The boundary layer was absent in the egress observation, which occurred in the Charon-facing hemisphere over a darker, ice-free area. The egress data indicated a warmer surface temperature and a lower atmospheric pressure (1.02 ± 0.07 Pa and 51.6 ± 3.8 K). Both occultation data revealed a strong temperature inversion at altitudes below 20 km, more pronounced during ingress (6.4 ± 0.9 K $km^{-1}$) compared to egress (3.4 ± 0.9 K $km^{-1}$). Above 30 km, the temperature profiles from both occultations converged, suggesting a possibly well-mixed temperature horizontally.

From the $N_2$ and $CH_4$ density distributions observed in ultraviolet solar occultation, a temperature of 65-68 K in the upper atmosphere of Pluto was derived. Combined with the radio occultation constraints in the lower atmosphere, the best-fit model suggests a temperature distribution on Pluto that begins with a thermal inversion near the surface, reaching a peak of about 107 K at 25 km, then decreasing with altitude to a broad minimum of 63 K around 400 km, and maintaining an isothermal range of 65-68 K above towards the exobase at around 1710 km. This observed upper atmosphere is significantly colder and more compact than the pre-New Horizons estimates.

$CH_4$ abundances, measured from 80 to 1200 km above the surface, indicated a surface mixing ratio of about 0.28–0.35%. The best $CH_4$ model points to a low eddy diffusion coefficient ranging from 550 to 4000 $cm^2$ $s^{-1}$ in the lower atmosphere, suggesting a homopause height of less than 12 km on Pluto. Ultraviolet occultation also revealed the vertical profiles of photochemical products $C_2H_2$, $C_2H_4$, and $C_2H_6$ in the middle and upper atmosphere. Local maxima for $C_2H_4$ at 410 km, $C_2H_2$ at 320 km, and a potential local maximum for $C_2H_6$ at 260 km suggest that their production regions are approximately 300-400 km. Occultation data also show local minima or inflection points for these hydrocarbons: $C_2H_4$ near 200 km, $C_2H_2$ near 170 km, and $C_2H_6$ between 170–200 km. Chemical models that include ice condensation and heterogeneous chemistry on aerosols around 200 km can account for the local depletion of hydrocarbons. HCN was detected via radio telescopes and supersaturated by many orders of magnitude in the upper atmosphere.

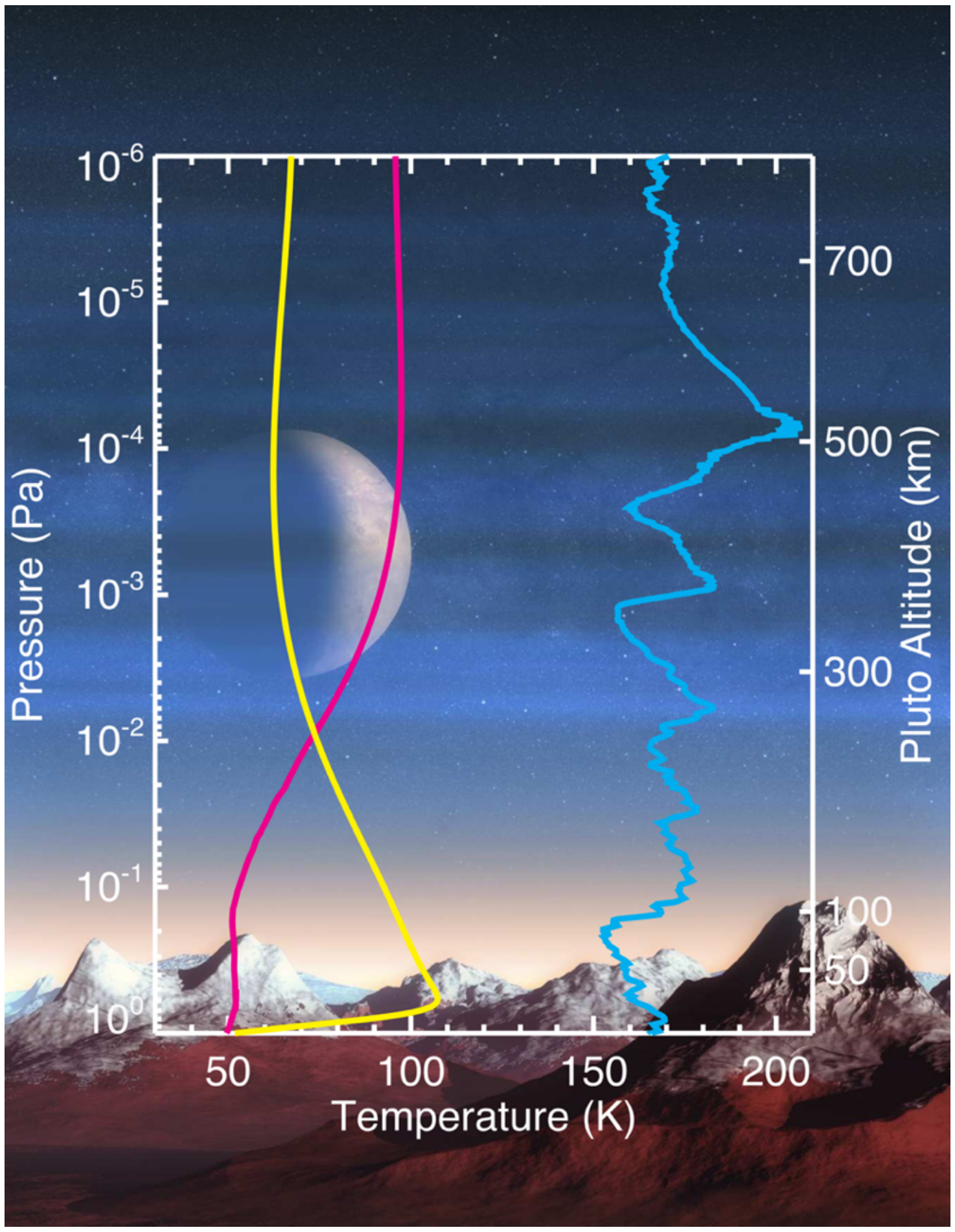

***Figure 6.*** *Temperature-pressure profiles of Pluto (yellow), Triton (magenta), and Titan's upper atmosphere (blue). The vertical axis uses Pluto's pressure and altitude grid. The Pluto data was from Zhang et al. (2017) and Young et al. (2018). The Triton data was provided by Darrell Strobel. The Titan data was from the Huygens Probe (Fulchignoni et al. 2005). The background image depicts an artistic impression of Pluto's surface featuring tall mountains and a dark surface in the Cthulhu Macula region, with Charon visible in the sky through Pluto's atmospheric hazes (image credit: Pockn Liu).*

To the first order, the chemistry in Pluto's $N_2$-$CH_4$-CO atmosphere is analogous to that of Titan. Photochemistry involving $N_2$ and $CH_4$ yields an abundance of hydrocarbons and nitriles, as detailed in the previous section on Titan. However, there are three important differences. First, Pluto receives less than one-tenth the sunlight that reaches Titan, making photons from the interstellar medium more important in the photochemistry. Second, Pluto lacks an ionosphere, with an upper limit on electron density of 1000 $cm^{-3}$, five times lower than that of Titan, likely due to significantly reduced photoionization, the absence of a nearby giant planet's magnetosphere, and high $CH_4$ abundance leading to rapid electron recombination. Consequently, ion-chemistry is less effective on Pluto than on Titan. Third, given the cold temperatures and abundant haze particles in Pluto's atmosphere (Figure 6), ice condensation and heterogeneous chemistry play crucial roles.

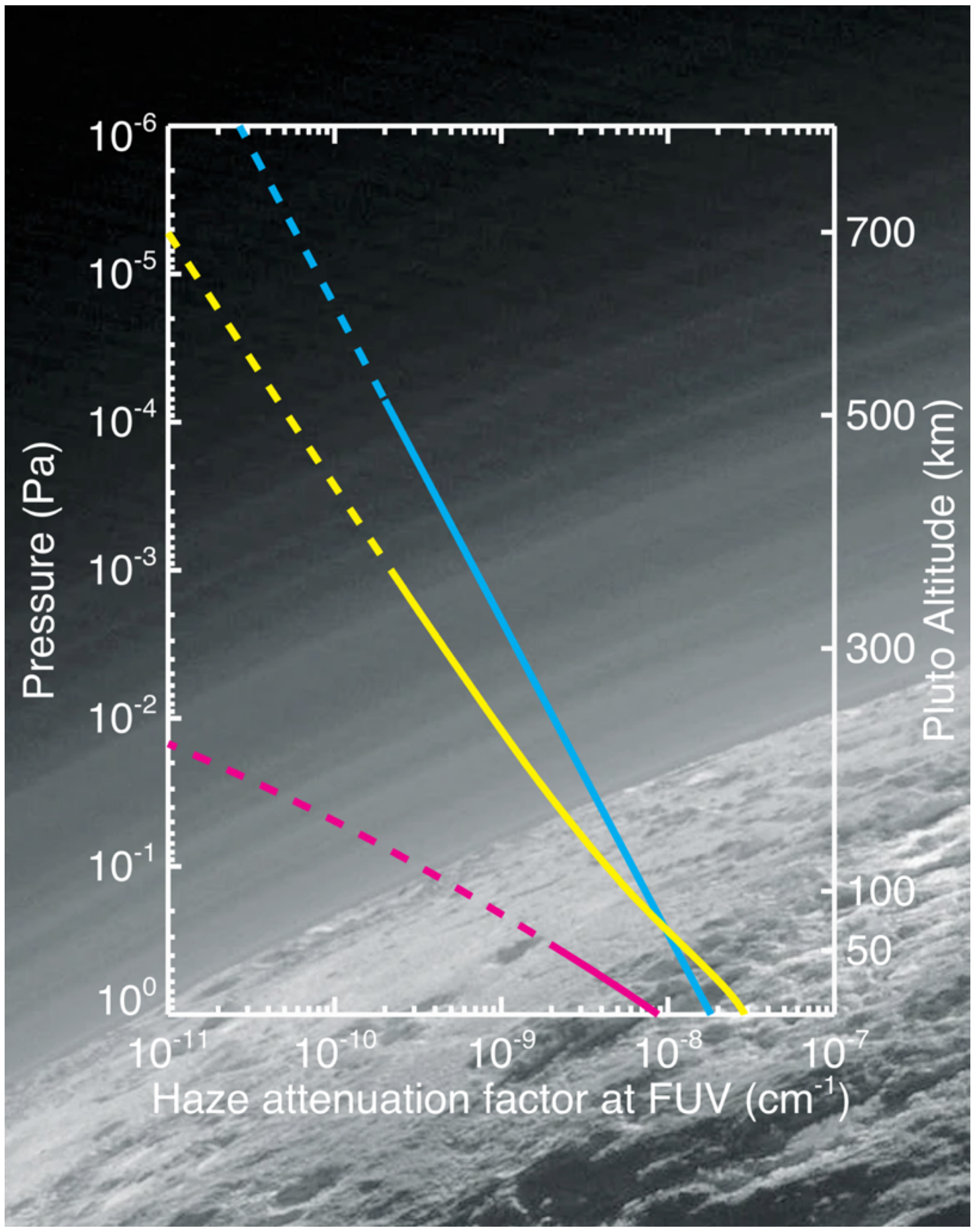

***Figure 7.*** *Vertical distribution of haze attenuation in the far-ultraviolet for Pluto (yellow), Triton (magenta), and Titan's upper atmosphere (blue), with extrapolated dashed lines. The vertical axis uses Pluto's pressure and altitude grid. The background image captures a near-sunset view of icy mountains and flat plains extending to Pluto's horizon, as captured by the New Horizons mission (NASA image: PIA19947), revealing more than a dozen layers of haze in Pluto's atmosphere. The Pluto data was based on New Horizons observations (Young et al. 2018). The Triton data was based on Voyager observations (Krasnopolsky et al. 1992). The Titan data was based on Cassini observations (Koskinen et al. 2011).*

Pluto is characterized by abundant hazes, with numerous distinct layers extending from near the surface to above 200 km, as clearly seen in New Horizons images (Figure 7). These hazes are globally distributed with a slightly higher concentration near the north pole. Approximately 20 distinct haze layers, each 1-3 km thick and typically spaced about 10 km apart, are observed up to an altitude of 200 km. UV occultation data further reveal that haze extinction extends above 350 km, with the extinction coefficient roughly proportional to the $N_2$ density. Intriguingly, the haze extinction profile on Pluto is comparable to that of Titan at equivalent pressures (Figure 7). The forward-scattering and bluish color of the haze particles suggest that they are fractal aggregates like the Titan haze, with a vertical scattering optical depth of about 0.004. Multi-wavelength, multi-phase-angle retrievals have revealed that the haze particles likely have a bi-modal distribution: a larger-size mode consisting of 1-micron two-dimensional aggregates with about 20 nm monomers, and a smaller-size mode of about 80 nm spheres. The concentrations of these two modes are approximately 0.3 $cm^{-3}$ for the aggregates and 10 $cm^{-3}$ for the spheres, with the total masses of the two modes being nearly equivalent. The origin of these two modes remains unclear.

If Pluto's haze is produced via a chemical pathway similar to that of Titan's haze, the required mass flux suggests that the conversion efficiency of $CH_4$ photolysis to haze particles is twice as large on Pluto as on Titan. Alternatively, it has been proposed that Pluto's hazes could form through direct ice condensation of photochemical products. The icy haze formation process begins with HCN ion nucleation and condensation in the upper atmosphere. As they settle, other gases (e.g., $C_3H_4$, $C_4H_2$, and $C_6H_6$) begin to condense onto the HCN ice cores, contributing about two-thirds of the haze mass flux, with a major contribution from $C_4H_2$. Below 400 km, collisions among particles lead to a transition from spherical growth due to gas deposition to aggregate growth due to collisions. Consequently, the fractal dimension of the particle changes from three to about two at around 300 km. The $C_2$ hydrocarbons ($C_2H_6$, $C_2H_4$, $C_2H_2$) play essential roles in particle growth through heterogeneous processes, contributing another one-third of the mass flux. These processes also result in the observed local abundance minimum for $C_2$ gases at around 200 km, as detected by New Horizons. As the particles further settle and pass through the temperature peak in the lower atmosphere, a thick coating could inhibit ice sublimation back, although the morphology such as fractal dimension might change. But this scenario does not reproduce the bi-modal particle distribution. The general prediction is that Pluto's hazes are a co-condensation ice mixture. Although co-condensation could also occur in Titan's stratosphere, the haze chemical composition in the two atmospheres—and consequently, their optical properties and impacts on the atmospheric radiative energy budget—would differ.

Pluto's atmosphere is primarily influenced by radiation below 700 km, while above this altitude, thermal conduction predominates. The unexpectedly cold temperatures observed above 400 km suggest that previous calculations may have overlooked a critical coolant. Atmospheric heating is largely attributed to solar absorption in the $CH_4$ near-infrared bands, where non-LTE effects are important. The atmosphere is cooled through non-LTE cooling by $CH_4$ and $C_2H_2$, as well as LTE rotational cooling by CO and HCN. While water vapor could significantly cool the atmosphere, the required concentrations would be supersaturated by eighteen orders of magnitude at those temperatures. Instead, the abundant aerosols in Pluto's atmosphere may provide a viable cooling mechanism. The particle cooling timescale in Pluto's thin atmosphere is much shorter than that for gas cooling or conduction, suggesting that gas and particles are likely in thermal equilibrium below 700 km. The large haze cooling leads to a shorter radiative timescale of the atmosphere, which is consistent with the unexpected latitudinal variations of Pluto's lower atmosphere observed from ALMA.

Given the largely unknown composition of Pluto's haze, the refractive indices are not well constrained. Nonetheless, radiation models across a broad parameter range consistently indicate that haze heating and cooling rates substantially exceed those of the gases. If haze cooling is sufficiently strong, the predicted thermal emission from the haze would make Pluto appear much brighter than its surface in mid-infrared wavelengths—a phenomenon later confirmed by observations from the James Webb Space Telescope (JWST). However, these JWST observations were predominantly sensitive to the lower atmospheric hazes. According to the described haze formation scenario, the composition of Pluto's haze could evolve due to co-condensation and ice coating as they settle. If haze radiation properties vary substantially between the middle and upper atmospheres, the unresolved issue of cold temperatures on Pluto remains.

In the lower atmosphere, heat transport was previously thought to be governed by thermal conduction and gas radiation. However, the dominance of haze radiation challenges this traditional view. If haze heating and cooling are robust, the radiative-equilibrium solution would not align with the observed thermal inversion in the lower 25 km, unless the haze is exceptionally bright near the surface. While a bright haze due to ice coating is plausible, it is more likely that heat mixing by atmospheric eddies and turbulence, rather than weak thermal conduction, dominates the heat transport, linking the warm atmosphere to the cold surface. The required eddy heat diffusivity is estimated at about 1000 $cm^2\ s^{-1}$, aligning with the chemical eddy diffusivity derived from chemical models and theoretical estimates from general circulation models of Pluto.

The atmospheric dynamics and general circulation patterns on Pluto remain loosely constrained by the New Horizons data. As Pluto is a moderate rotator with a rotation period of about 6.4 Earth days, the Coriolis force still plays a key role in shaping atmospheric circulation. The atmosphere exhibits a small spatial variation, attributed to the cold temperatures which extend the radiative timescale to between 1-10 Earth years, depending on the aerosol radiative properties. Pluto's surface has three key influences on the atmospheric flows. First, the cold surface underlying the warmer atmosphere creates strong static stability in the lower atmosphere, characterized by a pronounced thermal inversion and low dynamical mixing, with a homopause located below 12 km. Secondly, strong topographic features significantly modify local flows. On Pluto, local downslope katabatic winds can emerge due to the cooling of air along slopes, leading to denser downward flows, particularly near topographical landmarks like Sputnik Planitia. Thirdly, the condensation and sublimation of nitrogen ice on the surface generate substantial pressure gradients, which are the principal drivers of Pluto's atmospheric circulation. Winds flow from sunlit subliming regions to shaded condensing areas, with their direction influenced by the Coriolis effect. This sublimation-condensation dynamics also contributes to the formation of the boundary layer observed over Sputnik Planitia.

Dynamical simulations have confirmed that atmospheric flow on Pluto is predominantly governed by $N_2$ condensation and sublimation, as well as by the topography. However, models can yield either globally prograde (eastward) or retrograde (westward) zonal winds, crucially dependent on the distribution of nitrogen ice on Pluto's surface. Incorporating realistic topographic data and ice distribution from the 2015 New Horizons observations revealed that an intense near-surface circulation forms above Sputnik Planitia, characterized by an anticlockwise spiral flow and a western boundary current, similar to those observed on Earth and Mars (Figure 5). This near-surface circulation and a hemispheric temperature gradient facilitated meridional transport of sublimated $N_2$ from the northern summer to the southern winter hemisphere in 2015. The meridional winds are weak, with speeds less than 1 $m\ s^{-1}$, but due to the conservation of angular momentum, this cross-equatorial transport generated a retrograde wind of up to 10 $m\ s^{-1}$ over the equator.

Pluto's dynamic atmosphere is characterized by thermal tides, gravity waves, and barotropic waves, likely induced by orographic forcing, jet instabilities, and/or the sublimation-condensation cycles of inhomogeneously distributed surface ices. Fluctuations in temperature and density derived from stellar occultation light curves indicate the presence of these atmospheric waves. These waves exhibit vertical wavelengths ranging from 3-6 km in the lower atmosphere to 5-20 km in the upper atmosphere. Notably, these waves are thought to contribute to the approximately 20 distinct haze layers. Their typical spacing of 10 km over hundreds of kilometers indicates of planetary-scale wave activity. These layers appear more pronounced near the equator than at mid-latitudes, suggesting confinement to low latitudes. The absence of observed temporal variation in these layers suggests that the waves could be stationary, or that their periods are significantly longer than a few hours. Stationary inertia-gravity waves, excited by surface topography, might explain the observed haze layers, although a background wind speed of 2 m $s^{-1}$, required to account for the vertical spacing, appears inconsistent with the highly variable winds indicated by GCMs. While thermal tides can replicate the spacing, the associated temperature fluctuation of only 0.2 K is too low to account for the distinct haze layers.

Pluto's atmosphere also demonstrates strong temporal variations across different scales. Locally, sublimation-condensation flows operate on a diurnal scale, with winds flowing out of the Sputnik Planitia basin in the afternoon and returning at night and in the early morning. Seasonal variations are strong due to Pluto's high obliquity of 119.6 degrees and its highly eccentric orbit with an eccentricity of 0.25. The surface pressure varies by orders of magnitude within a single Pluto year (248 Earth years), a trend observed over recent decades. Despite this, even at aphelion (approximately 49.9 AU from the Sun), the atmosphere is unlikely to collapse due to the sublimation from equatorial $N_2$ sources in Sputnik Planitia and the high thermal inertia of the subsurface. It is anticipated that meters of $N_2$ ice may be transported globally over the seasons, though the detailed transport patterns and resultant ice distribution remain unclear. Various factors influence these processes, including general circulation, ice mixture thermodynamics, surface and subsurface thermal inertia, and the largely uncharted ice distribution in the southern hemisphere beyond 33 degrees, which was not imaged during the New Horizons flyby. Over Milankovitch timescales (~3 Myrs), Pluto's climate could undergo dramatic changes, with local accumulation or erosion of meter- to kilometer-thick layers of volatile ice.

The origins of Pluto's volatiles remain elusive. Its nitrogen could have been accreted as protosolar $N_2$ or derived from ammonia. Unlike Titan, observational data such as $^{15}N/^{14}N$ and $^{36}Ar/^{14}N$ ratios, which could help differentiate these sources, are not available for Pluto. Additionally, Pluto may have inherited substantial organics that contribute to its volatile inventory. Observations of the lower limit of $^{14}N/^{15}N$ from atmospheric HCN suggest a protosolar $N_2$ origin, although the fractionation process for HCN is highly uncertain. The very low $CO/N_2$ ratio at Pluto's surface might suggest an origin of nitrogen as ammonia, though it is possible that the majority of CO ice is buried beneath the surface.

Regarding volatile loss, Pluto's atmosphere primarily loses material through irreversible escape mechanisms, including photodissociation and photoionization. Photochemical processes account for more than 90% of the total $CH_4$ loss rate and about half of the $N_2$ loss, while losses due to the solar wind are negligible. Owing to the cold conditions in the upper atmosphere, the escape rates are estimated at (3–8) × $10^{22}$ $N_2$ $s^{-1}$ and (4–8) × $10^{25}$ $CH_4$ $s^{-1}$ at the exobase. The atmospheric residence times are approximately 720 Earth years for $CH_4$ and about 4.4 million years for $N_2$. If these loss rates have remained constant, the current $N_2$ inventory in Sputnik Planitia would be sufficient to sustain $N_2$ loss for over 100 billion years. The $CH_4$ ice inventory, including all bladed terrain, might also be adequate, as $CH_4$ loss is likely diffusion-limited. However, these loss rates are probably not constant, due to drastic climate changes on Pluto, the secular evolution of solar flux, and orbital changes during the giant planet instability era in the early Solar System. Moreover, large uncertainties arise from the largely unconstrained haze chemistry and opacity in Pluto's atmosphere, which could fundamentally influence the temperature and loss rates in the upper atmosphere over Pluto's geological history.

## 5. Triton

Triton, Neptune's largest moon, was discovered shortly after Neptune itself in 1846. However, our understanding of Triton and its atmosphere remained limited until the Voyager 2 flyby in 1989. Triton shares many similarities with Pluto; it is about 20% larger than Pluto, with a diameter of 1353 km compared to Pluto's 1188 km. The semi-major axis of Neptune's orbit is about 30 AU, approximately 30% closer to the Sun than Pluto's orbit at 39.5 AU, though its orbital eccentricity (0.0087) is much less than Pluto's (0.25). Triton's rotation period is approximately 5.9 days, close to Pluto's 6.4 days. Both bodies exhibit similar surface pressures and temperatures, with thin but hazy atmospheres primarily composed of $N_2$-$CH_4$-CO in equilibrium with their surface ices. Despite these similarities, detailed observations from Voyager on Triton and New Horizons on Pluto highlight distinct differences between these two celestial siblings.

During the Voyager flyby, only 40% of Triton's surface was imaged, as the northern hemisphere poleward of approximately 45°N was obscured by polar night (Figure 8). Triton's surface geology significantly differs from Pluto's; it is much flatter, with topographic variations typically less than a few hundred meters, in contrast to Pluto's mountainous regions which can rise several kilometers. Furthermore, Triton exhibits less contrast in albedo and color than Pluto. Its southern hemisphere is brighter than the northern hemisphere. Triton does not feature large, dark, and volatile-free terrains such as the Cthulhu Regio on Pluto. Triton's surface is covered by 55% volatile ice, primarily beta-phase $N_2$, with 0.1% $CH_4$ and 0.05% CO in the mixture. The remaining 45% comprises discrete, non-volatile ice units, including 10-20% $CO_2$ (absent on Pluto) and 25-35% $H_2O$. The estimated surface age of 5-50 million years suggests active resurfacing processes, possibly due to cryovolcanism. Similar to Pluto, the distribution of surface ice plays a crucial role in shaping atmospheric dynamics through sublimation-condensation flows. However, a key difference lies in the nitrogen reservoirs: on Pluto, the nitrogen ices are predominantly located in the Sputnik Planitia in the equatorial region, whereas on Triton, the largest ice reservoir is thought to be the extensive bright southern cap. It is also hypothesized that there could be another ice cap near the unimaged north pole during the Voyager flyby.

The surface pressure on Triton, derived from Voyager's radio occultation, is between 1.2-1.4 Pa, and the surface temperature is established at 37.5 ± 0.5 K, assuming vapor pressure equilibrium between the $N_2$ gas and ice. This measurement aligns with infrared data, which suggests a temperature of 41 ± 5 K (assuming unity surface emissivity), and is consistent with ground-based spectroscopy of the $N_2$ ice band at 2.16 µm. Triton possesses a troposphere, evidenced by the presence of condensation clouds and the observation of plume heights within the lowest 8 km. Sublimation-driven winds facilitate forced convection, resulting in a negative temperature gradient that follows the moist adiabat of -0.1 K km$^{-1}$ from the surface to the tropopause at approximately 8 km.

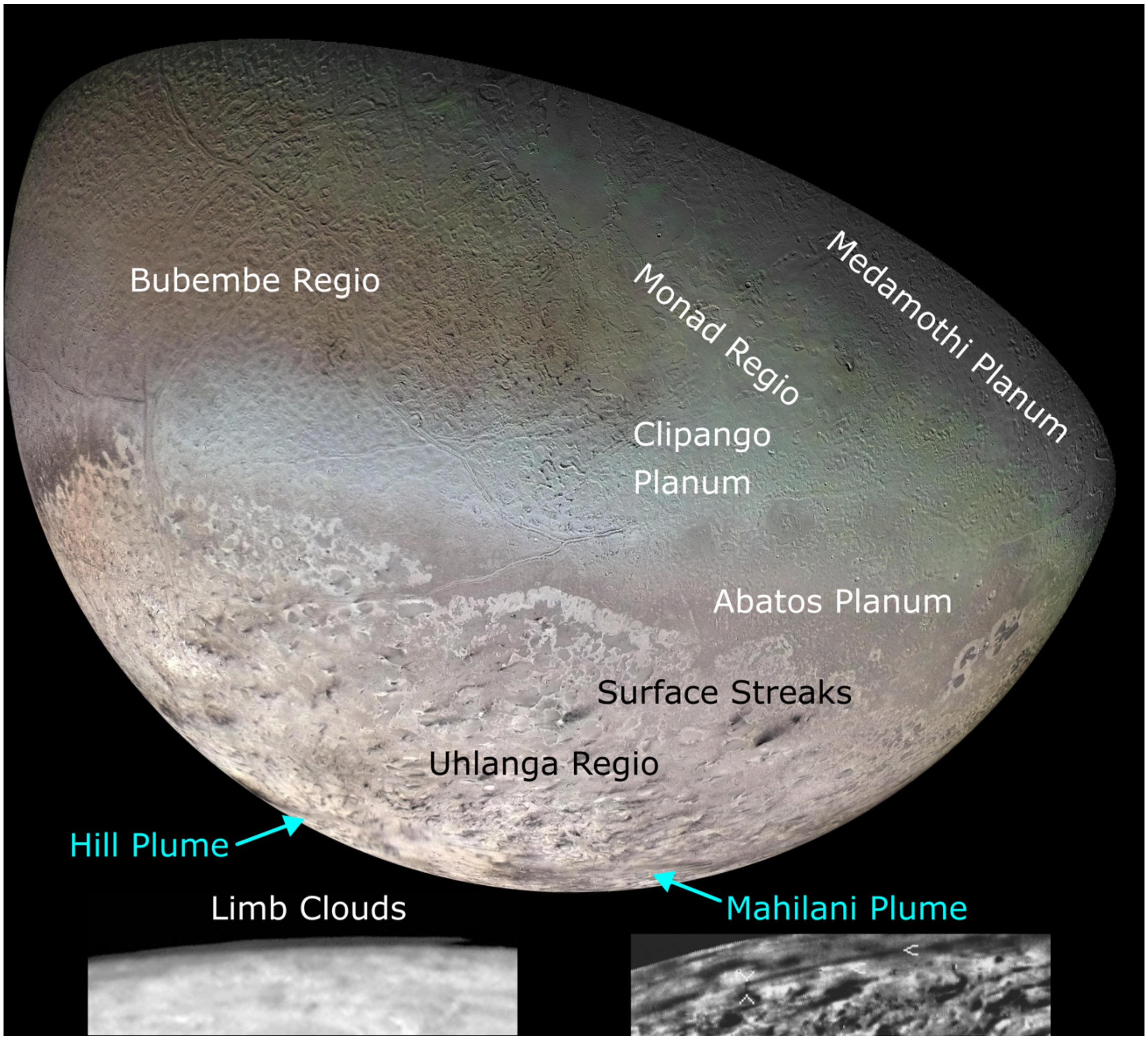

***Figure 8.*** *Geologic and atmospheric features observed on Triton during the Voyager 2 flyby. Many dark surface streaks are prominent in the southern hemisphere of the Triton image (credit: PIA00317). Additionally, two active plumes were captured, with a close-up view of the Mahilani Plume on the lower right (NASA image: PIA14448). A subset on the lower left highlights the condensable nitrogen clouds observed on the limb of Triton (NASA image: PIA02203).*

UV occultation data have determined the $N_2$ density distribution, suggesting an isothermal upper thermosphere, significantly warmer at about 102 ± 3 K from 475 to 675 km (Figure 6). The atomic nitrogen profile from 170 to 600 km also indicates a temperature of 100 ± 7 K at 400 km. The consistency of results at both ingress and egress observations suggests that the atmosphere may be horizontally homogeneous. However, the temperature profile between the surface and approximately 150 km was not probed by Voyager but was later derived from ground-based stellar occultation. Triton's atmosphere exhibits near isothermal conditions at around 50 K in the 25-50 km region, above which the thermal structure shows a steep, positive temperature gradient up to about 150 km, indicating a downward heat flux from the upper atmosphere.

Besides nitrogen, UV occultation also constrained the $CH_4$ profile below 80 km. In contrast to Pluto and Titan, where $CH_4$ is well mixed in the lower atmosphere, Triton's $CH_4$ mole fraction is about 0.02% (200 ppm) near the surface but decreases rapidly with altitude due to photolysis by Lyman-α photons from the sun and the interstellar medium. The dissociated $CH_4$ cannot be replenished by upward diffusive transport to maintain a constant mixing ratio with altitude, putting a constraint on the vertical eddy diffusivity between 300-500 $cm^2\ s^{-1}$ from 10-50 km. The CO mixing ratio, inferred from surface-ice spectra, is approximately 0.015% (150 ppm). Ground-based submillimeter observations in 2016 detected about 30 ppm HCN and 230 ppm CO on Triton. Surprisingly, Voyager detected a robust ionosphere on Triton with peak electron densities of 2-5 $\times 10^4\ cm^{-3}$ at 340-350 km, much higher than that observed on other giant planet moons, including Titan. Additionally, Voyager recorded airglow emissions from $N_2$ and $N^+$, as well as H Lyman-α emissions, with spectra similar to those observed on Titan.

Even though Triton's atmosphere is composed of $N_2$, $CH_4$, and CO, its chemistry distinctly differs from that of Titan and Pluto. Due to the depletion of $CH_4$ in the upper atmosphere, typical Titan-like hydrocarbon photochemistry is confined to below 200 km on Triton. The products, including $C_2H_2$, $C_2H_4$, $C_2H_6$, and other hydrocarbons and nitriles, condense in the cold atmosphere, and their abundances were too low to be detected by Voyager. In the upper atmosphere, abundant CO could provide a significant source of carbon and oxygen. Atomic species and their ions are far more prevalent on Triton than on Titan and Pluto. The major ion in the ionosphere is likely $C^+$, with $N^+$ being abundant above the peak, and $HCO^+$ below 250 km. Although some models can generate the ionosphere using only solar EUV photons, it is believed that precipitation of high-energy particles from Neptune's magnetosphere plays a crucial and possibly dominant role in ionization. Moreover, Neptune's magnetosphere is enriched by the thermal escape of neutral H, $H_2$, and N from Triton's exobase at around 900 km, with escape rates of hydrogen and nitrogen estimated at $10^{26}\ s^{-1}$ and $10^{25}\ s^{-1}$, respectively.

Neptune's particle precipitation not only greatly alters the chemical makeup of Triton's atmosphere but also affects its thermal structure. Compared to Pluto, Triton's lower atmosphere is colder, whereas the upper atmosphere is considerably warmer. A depletion of $CH_4$ implies that Triton has much less $CH_4$ heating than Pluto. However, Triton's upper atmosphere experiences intense heating due to solar EUV flux and Neptune's magnetospheric power dissipation, facilitated by direct and indirect $N_2$ recombination reactions. This substantial energy influx from the upper atmosphere diffuses thermally into the lower layers, which are cooled by CO rotational lines, similar to processes observed on Pluto. In the lowermost 8 km, where a possible troposphere exists with a negative temperature gradient, eddy diffusion is hypothesized to vertically mix heat, similar to the near-surface layers on Pluto. The required eddy diffusivity at 8 km is about 300 $cm^2\ s^{-1}$, aligning with the chemistry eddy diffusivity needed to explain the $CH_4$ vertical profile.

Thermal structure models for Triton that incorporate $CH_4$ heating, CO cooling, magnetic electron heating, and thermal conduction produce a temperature profile largely consistent with observations, except for a discrepancy: the models predict a strong positive temperature gradient rather than the observed near isothermal conditions between 10-50 km. This discrepancy suggests an efficient cooling mechanism is missing below 50 km. Given the role of haze heating and cooling observed on Pluto, the lower atmospheric haze on Triton may be an important coolant to consider.

Due to the rapid decrease of $CH_4$ with altitude on Triton, the photochemical hazes are concentrated at significantly lower altitudes (<30 km), unlike the extended haze layers observed on Pluto (Figure 7). Low and high phase angle images from Voyager illustrate that haze particles are globally distributed. The vertical profile of UV (0.15 micron) attenuation (Figure 7) indicates that haze extinction on Triton is only about 2-3 times lower than that on Pluto in the near-surface levels but drops much more rapidly with altitude. The vertical optical depth ranges from 0.01-0.02 in the UV to 0.003-0.004 in the visible wavelengths. The mean particle radius is estimated to be between 0.15-0.2 microns. Triton's haze appears much brighter than that of Titan, as the absorbing "tholin-like" Titan haze cannot explain the UV-to-visible opacity wavelength dependence. Similar to the hazes on Pluto, Triton's haze may also consist of condensable ices of hydrocarbons and nitriles near the surface. Microphysical models suggest that particle formation begins with the nucleation of HCN in the upper atmosphere, yet efficient growth to fractal aggregates occurs in the lower atmosphere where $CH_4$ photochemical products are abundant. The cold temperatures allow supersaturated hydrocarbons to condense directly onto aerosols, with $C_2H_4$ ice predicted to be the predominant component, followed by $C_2H_6$ ice.

These haze particles may also act as condensation nuclei for the observed bright, patchy, nitrogen clouds near the surface, predominantly found at middle and high southern latitudes, covering approximately 37% of Triton poleward of 30°S but only 6% in equatorial and northern regions. During the crescent phase, Voyager detected six long bright cloud streaks approximately 10 km wide and between 77-254 km in length, located at 1-3 km above the surface. The cloud particles' scale height suggests they are vertically well mixed. These clouds could be naturally produced by supersaturated nitrogen air due to adiabatic cooling in sublimation-driven convection. Another potential origin is surface vents releasing high-pressure nitrogen from the subsurface, similar to the mechanism producing Triton's distinctive plumes.

Geyser-like plumes have been observed over Triton's south polar cap (Figure 8). During the Voyager flyby, at least two active plumes were identified near the subsolar latitude at 50–57°S, erupting columns of dark materials ranging in size from 2-20 km. The ejected gas and dust rise vertically up to the tropopause at around 8 km and spread horizontally, forming long tails approximately 150 km in length. The estimated mass flux includes a few kg $s^{-1}$ of fine dark particles and/or condensed volatiles and up to several hundred kg $s^{-1}$ of vapor. Additionally, more than 120 fan-shaped dark streaks, ranging from 4 to over 100 km in length, were identified on the surface between latitudes of 15° and 45°S. These streaks, similar to the "wind tails" observed on Mars, are likely deposits from eruptions that were inactive during the Voyager flyby.

The origin of Triton's plumes remains enigmatic, with several proposed mechanisms that include both exogenic influences—similar to solar-driven jets on Mars—and endogenic processes like the geysers on Saturn's moon Enceladus. One exogenic theory, the "solid-state greenhouse" mechanism, suggests that solar heating of thermally insulated nitrogen over a dark subsurface could melt the base, leading to pressurized escape of ice and dark particles into the atmosphere. Calculations indicate that a modest two-degree warming is sufficient to power these geysers. Once the materials ascend above the surface, vapor condensation during the rising plume could release sufficient latent heat to accelerate the upwelling, potentially causing it to overshoot the tropopause. However, the absence of similar plume activity on Pluto's nitrogen ice surfaces, as observed by New Horizons, might cast doubt on the exogenic mechanism. On the other hand, endogenic mechanisms might be at work including basal melting of a thick nitrogen polar cap due to geothermal heating, followed by melt eruption, or cryovolcanic outgassing from Triton's deep interior.

Wind directions on Triton can be inferred from the orientation of dark surface streaks, crescent cloud streaks, and plume tails. The primary direction of dark surface streaks range between 40° and 80°, measured clockwise from north, with an average orientation in the northeast quadrant at 59°. This suggests that near-surface winds predominantly blow toward the northeast. Cloud streaks indicate eastward movement at the 1-3 km level with speeds around 13 m $s^{-1}$. Rapidly rising plumes, which are almost vertical initially, do not exhibit wind effects until reaching the tropopause at about 8 km, where plume tail directions indicate westward winds. One observed plume tail expanded from 90 to over 150 km in just 90 minutes, implying a westward wind speed of about 15 m $s^{-1}$at about 8 km.

A general understanding of Triton's atmospheric dynamics during the Voyager era is as follows. The dynamics are primarily driven by the sublimation of nitrogen from the southern polar cap. An Ekman boundary layer, influenced by surface friction and the Coriolis effect, forms near the surface. Nitrogen is transported toward the northern winter hemisphere with a northeast wind averaging 5 m $s^{-1}$. Above 1 km, winds circulate to the east, forming an anticyclonic vortex over the south pole with speeds ranging from 10 to 20 m $s^{-1}$. At higher altitudes, the temperature gradient from the darker, warmer equator to the brighter, colder south pole generates a strong thermal wind. Given Triton's retrograde rotation, this thermal wind in the southern hemisphere weakens the eastward flow with altitude, eventually reversing to a westward direction above 3-5 km. However, the specifics of Triton's circulation pattern remain largely undefined, as no comprehensive GCM has yet been applied to study Triton's atmosphere.

Triton exhibits a more complex seasonal behavior than Pluto and Titan, driven by its unique orbital dynamics. As a synchronously rotating moon in retrograde orbit around Neptune, Triton's axial tilt of 50.4° (or 129.6°) relative to Neptune's orbital plane, combined with a short orbital precession period of about 688 years, results in a complex movement of the subsolar latitude, which oscillates between 5°, 20°, and 50° during the summer solstice. Similar to Pluto, models exploring volatile transport on Triton have considered both short and long-term changes but are hindered by uncertainties such as global ice distribution, internal heat flux, thermal inertia, albedo, emissivity, and atmospheric dynamics. Generally, these models predict that Triton's atmosphere is unlikely to collapse. Despite north-south asymmetry in surface properties, a small permanent polar cap could persist in the northern pole if the internal heat flux is low. Triton's poles act as cold traps for volatile ices, and the polar caps extend to lower latitudes either through viscous spreading or the accumulation of thin seasonal deposits.

Since Voyager's visit, the seasonal behavior of Triton has been driven by the sublimation of $N_2$ in the southern hemisphere and condensation in the northern hemisphere. Surface pressure changes on Triton have been tracked through ground-based observations during stellar occultations. It was observed that surface pressure peaked at approximately 2 Pa around the year 2000, coinciding with an extreme southern summer solstice—a phenomenon that occurs once every 650 years. However, observations from 2017 to 2022 show a roughly consistent pressure, differing from model predictions of a steady pressure decline from 2005 to 2060. Continuous monitoring of Triton, or even better, a revisit by another space mission, would significantly enhance our understanding of its climate dynamics.

As a retrograde rotator around Neptune, Triton is likely a captured Kuiper Belt object, indicating that its composition may not reflect the original circumplanetary disk of Neptune. The surface ices on Triton can be explained by solar nebula condensation models or cometary abundances. During its capture, Triton might have lost any possible original binary companion it had and likely underwent complete melting due to intense tidal heating. This massive atmosphere during Triton's molten phase could have led to substantial chemical differentiation between the atmospheric species and the underlying ocean, potentially explaining the low $CO/N_2$ and $CO/CO_2$ ratios observed in Triton's current surface ices. However, like Pluto, the origin of Triton's volatiles remains elusive due to the lack of isotopic measurements and data on noble gases such as argon.

## 6. Conclusions

Every atmosphere in the Solar System is unique. Among all eleven atmospheres, those of Io, Triton, and Titan stand out due to the strong influence of the magnetospheric particles from their host planets. In terms of atmospheric chemistry, Io's atmosphere bears similarities to the sulfur chemistry found in the middle atmospheres of Venus and the stratosphere of Earth. Conversely, the hydrocarbon chemistry in Titan, Triton, and Pluto resembles the reduced chemistry in the hydrogen-dominated atmospheres of the four giant planets, which also contain many hydrocarbons, nitriles, and photochemical hazes.

Broadly speaking, the Solar System atmospheres can be categorized into several climate regimes. The "Jovian regime" includes the four giant planets, where fast-rotating, hydrogen-dominated atmospheres extend deep into the interior without a surface. The "terrestrial regime" includes Earth, Venus, and Titan, characterized by their thick atmospheres and interactions with the surface. These three atmospheres all feature some form of hydrological cycle: the sulfuric acid cycle on Venus, the methane cycle on Titan, and the water cycle on Earth. Titan's rotation period (16 days) lies between that of Venus (>200 Earth days) and Earth (1 day), positioning the Coriolis effect between these two extremes. Titan also provides an example of a climate where moisture is not cold trapped in the lower atmosphere, offering insights into moist or runaway greenhouse scenarios in the terrestrial regime.

The "condensable regime" includes Mars, Io, Triton, and Pluto, where the main atmospheric constituents—such as $CO_2$ on Mars, $SO_2$ on Io, and $N_2/CH_4/CO$ on Triton and Pluto—can condense. In these atmospheres, temperature, pressure, and dynamics are substantially affected by the sublimation and condensation of surface ices. Notably, because the atmospheres in the condensable regime are usually thin, aerosols such as Martian dust, Io's volcanic dust, and hazes on Triton and Pluto play significant roles in the atmospheric energy balance and heterogeneous chemistry. The Kuiper belt object Eris might also fall into this regime if it has an atmosphere. $N_2$ and $CH_4$ ices have been detected on Eris, suggesting that it could have a local or global atmosphere with a pressure on the order of nanobars, similar to that on Io. Future stellar occultations should reveal whether Eris has the twelfth atmosphere in the Solar System.

There could arguably be a fourth "exosphere regime," which includes the exospheres on Mercury, Earth's moon, the asteroid Ceres, Jupiter's moons Ganymede, Callisto, and Europa, as well as Saturn's moon Enceladus. These atmospheres are characterized by their non-collisional nature and strong interactions with the space environment. For the icy moons, the exospheres are closely connected with surface ices and potentially the internal oceans beneath the ice shells, through processes like ice-ocean convection and ejected plumes, as observed on Enceladus and possibly on Europa. These atmospheric features are crucial for understanding the astrobiological potential of the icy moons, though this topic extends beyond the scope of this article.

**References**

Achterberg, R.K., 2023. Temporal evolution of titan's stratospheric temperatures and trace gases from a two-dimensional retrieval of cassini composite infrared spectrometer data 4, 140. https://doi.org/10.3847/PSJ/acebea

Bertrand, T., Forget, F., Schmitt, B., White, O.L., Grundy, W.M., 2020. Equatorial mountains on Pluto are covered by methane frosts resulting from a unique atmospheric process. Nature Communications 11, 5056. https://doi.org/10.1038/s41467-020-18845-3

de Pater, I., Goldstein, D., Lellouch, E., 2023. The plumes and atmosphere of io, in: Lopes, R.M.C., de Kleer, K., Keane, J.T. (Eds.), Io: A New View of Jupiter's Moon, Astrophysics and Space Science Library. pp. 233–290. https://doi.org/10.1007/978-3-031-25670-7_8

Emran, A., Dalle Ore, C.M., Ahrens, C.J., Khan, M.K.H., Chevrier, V.F., Cruikshank, D.P., 2023. Pluto's surface mapping using unsupervised learning from near-infrared observations of LEISA/Ralph 4, 15. https://doi.org/10.3847/PSJ/acb0cc

Fulchignoni, M., Ferri, F., Angrilli, F., Ball, A.J., Bar-Nun, A., Barucci, M.A., Bettanini, C., Bianchini, G., Borucki, W., Colombatti, G., Coradini, M., Coustenis, A., Debei, S., Falkner, P., Fanti, G., Flamini, E., Gaborit, V., Grard, R., Hamelin, M., Harri, A.M., Hathi, B., Jernej, I., Leese, M.R., Lehto, A., Lion Stoppato, P.F., López-Moreno, J.J., Mäkinen, T., McDonnell, J.A.M., McKay, C.P., Molina-Cuberos, G., Neubauer, F.M., Pirronello, V., Rodrigo, R., Saggin, B.,

Schwingenschuh, K., Seiff, A., Simões, F., Svedhem, H., Tokano, T., Towner, M.C., Trautner, R., Withers, P., Zarnecki, J.C., 2005. In situ measurements of the physical characteristics of Titan's environment 438, 785–791. https://doi.org/10.1038/nature04314

Gladstone, G.R., Young, L.A., 2019. New horizons observations of the atmosphere of pluto. Annual Review of Earth and Planetary Sciences 47, 119–140. https://doi.org/10.1146/annurev-earth-053018-060128

Hayes, A.G., Lorenz, R.D., Lunine, J.I., 2018. A post-Cassini view of Titan's methane-based hydrologic cycle. Nature Geoscience 11, 306–313. https://doi.org/10.1038/s41561-018-0103-y

Hörst, S.M., 2017. Titan's atmosphere and climate. Journal of Geophysical Research (Planets) 122, 432–482. https://doi.org/10.1002/2016JE005240

Koskinen, T.T., Yelle, R.V., Snowden, D.S., Lavvas, P., Sandel, B.R., Capalbo, F.J., Benilan, Y., West, R.A., 2011. The mesosphere and lower thermosphere of Titan revealed by Cassini/UVIS stellar occultations 216, 507–534. https://doi.org/10.1016/j.icarus.2011.09.022

Krasnopolsky, V.A., Sandel, B.R., Herbert, F., 1992. Properties of haze in the atmosphere of Triton 97, 11695–11700. https://doi.org/10.1029/92JE00945

Le Mouélic, S., Rodriguez, S., Robidel, R., Rousseau, B., Seignovert, B., Sotin, C., Barnes, J.W., Brown, R.H., Baines, K.H., Buratti, B.J., Clark, R.N., Nicholson, P.D., Rannou, P., Cornet, T., 2018. Mapping polar atmospheric features on Titan with VIMS: From the dissipation of the northern cloud to the onset of a southern polar vortex 311, 371–383. https://doi.org/10.1016/j.icarus.2018.04.028

Nixon, C.A., 2024. The composition and chemistry of titan's atmosphere. ACS Earth and Space Chemistry 8, 406–456. https://doi.org/10.1021/acsearthspacechem.2c00041

Strobel, D.F., Cui, J., 2014. Titan's upper atmosphere/exosphere, escape processes, and rates, in: Titan. p. 355. https://doi.org/10.1017/CBO9780511667398.013

Teanby, N.A., de Kok, R., Irwin, P.G.J., Osprey, S., Vinatier, S., Gierasch, P.J., Read, P.L., Flasar, F.M., Conrath, B.J., Achterberg, R.K., Bézard, B., Nixon, C.A., Calcutt, S.B., 2008. Titan's winter polar vortex structure revealed by chemical tracers. Journal of Geophysical Research (Planets) 113, E12003. https://doi.org/10.1029/2008JE003218

Bertrand, T., Forget, F., Schmitt, B., White, O.L. and Grundy, W.M., 2020. Equatorial mountains on Pluto are covered by methane frosts resulting from a unique atmospheric process. Nature communications, 11(1), p.5056.

Vuitton, V., Yelle, R.V., Klippenstein, S.J., Hörst, S.M., Lavvas, P., 2019. Simulating the density of organic species in the atmosphere of Titan with a coupled ion-neutral photochemical model 324, 120–197. https://doi.org/10.1016/j.icarus.2018.06.013

Young, L.A., Kammer, J.A., Steffl, A.J., Gladstone, G.R., Summers, M.E., Strobel, D.F., Hinson, D.P., Stern, S.A., Weaver, H.A., Olkin, C.B., Ennico, K., McComas, D.J., Cheng, A.F., Gao, P., Lavvas, P., Linscott, I.R., Wong, M.L., Yung, Y.L., Cunningham, N., Davis, M., Parker, J.Wm., Schindhelm, R., Siegmund, O.H.W., Stone, J., Retherford, K., Versteeg, M., 2018. Structure and composition of Pluto's atmosphere from the New Horizons solar ultraviolet occultation 300, 174–199. https://doi.org/10.1016/j.icarus.2017.09.006

Zhang, X., Strobel, D.F., Imanaka, H., 2017. Haze heats Pluto's atmosphere yet explains its cold temperature 551, 352–355. https://doi.org/10.1038/nature24465

## Further Reading

Bagenal, F., Dols, V., 2023. Space environment of io, in: Lopes, R.M.C., de Kleer, K., Keane, J.T. (Eds.), Io: A New View of Jupiter's Moon, Astrophysics and Space Science Library. pp. 291–322. https://doi.org/10.1007/978-3-031-25670-7_9

Cheng, A.F., Summers, M.E., Gladstone, G.R., Strobel, D.F., Young, L.A., Lavvas, P., Kammer, J.A., Lisse, C.M., Parker, A.H., Young, E.F., Stern, S.A., Weaver, H.A., Olkin, C.B., Ennico, K., 2017. Haze in pluto's atmosphere 290, 112–133. https://doi.org/10.1016/j.icarus.2017.02.024

Fan, S., Gao, P., Zhang, X., Adams, D.J., Kutsop, N.W., Bierson, C.J., Liu, C., Yang, J., Young, L.A., Cheng, A.F., Yung, Y.L., 2022. A bimodal distribution of haze in Pluto's atmosphere. Nature Communications 13, 240. https://doi.org/10.1038/s41467-021-27811-6

Flasar, F.M., Achterberg, R.K., Schinder, P.J., 2014. Thermal structure of Titan's troposphere and middle atmosphere, in: Titan. p. 102. https://doi.org/10.1017/CBO9780511667398.006

Forget, F., Bertrand, T., Hinson, D., Toigo, A., 2021. Dynamics of pluto's atmosphere, in: Stern, S.A., Moore, J.M., Grundy, W.M., Young, L.A., Binzel, R.P. (Eds.), The Pluto System after New Horizons. pp. 297–319. https://doi.org/10.2458/azu_uapress_9780816540945-ch013.

Griffith, C.A., Rafkin, S., Rannou, P., McKay, C.P., 2014. Storms, clouds, and weather, in: Titan. p. 190. https://doi.org/10.1017/CBO9780511667398.009

Hansen, C.J., Castillo-Rogez, J., Grundy, W., Hofgartner, J.D., Martin, E.S., Mitchell, K., Nimmo, F., Nordheim, T.A., Paty, C., Quick, L.C., Roberts, J.H., Runyon, K., Schenk, P., Stern, A., Umurhan, O., 2021. Triton: Fascinating moon, likely ocean world, compelling destination! 2, 137. https://doi.org/10.3847/PSJ/abffd2

Hofgartner, J.D., Buratti, B.J., Hayne, P.O., Young, L.A., 2019. Ongoing resurfacing of KBO Eris by volatile transport in local, collisional, sublimation atmosphere regime 334, 52–61. https://doi.org/10.1016/j.icarus.2018.10.028

Ingersoll, A.P., 1990. Dynamics of triton's atmosphere 344, 315–317. https://doi.org/10.1038/344315a0

Kirk, R.L., Soderblom, L.A., Brown, R.H., Kieffer, S.W., Kargel, J.S., 1995. Triton's plumes: discovery, characteristics, and models., in: Neptune and Triton. pp. 949–989.

Lavvas, P., Lellouch, E., Strobel, D.F., Gurwell, M.A., Cheng, A.F., Young, L.A., Gladstone, G.R., 2021. A major ice component in Pluto's haze. Nature Astronomy 5, 289–297. https://doi.org/10.1038/s41550-020-01270-3

Lebonnois, S., Flasar, F.M., Tokano, T., Newman, C.E., 2014. The general circulation of Titan's lower and middle atmosphere, in: Titan. p. 122. https://doi.org/10.1017/CBO9780511667398.007

Lellouch, E., Gurwell, M.A., Moreno, R., Vinatier, S., Strobel, D.F., Moullet, A., Butler, B., Lara, L., Hidayat, T., Villard, E., 2019. An intense thermospheric jet on Titan. Nature Astronomy 3, 614–619. https://doi.org/10.1038/s41550-019-0749-4

Lellouch, E., McGrath, M.A., Jessup, K.L., 2007. Io's atmosphere, in: Lopes, R.M.C., Spencer, J.R. (Eds.), Io after Galileo: A New View of Jupiter's Volcanic Moon. p. 231. https://doi.org/10.1007/978-3-540-48841-5_10

Lorenz, R.D., Brown, M.E., Flasar, F.M., 2010. Seasonal change on titan, in: Brown, R.H., Lebreton, J.-P., Waite, J.H. (Eds.), Titan from Cassini-Huygens. p. 353. https://doi.org/10.1007/978-1-4020-9215-2_14

McKinnon, W.B., Glein, C.R., Rhoden, A.R., 2019. Formation, composition, and history of the pluto system: A post-new-horizons synthesis, in: LPI Editorial Board (Ed.), Pluto System after New Horizons, LPI Contributions. p. 7067.

McKinnon, W.B., Lunine, J.I., Banfield, D., 1995. Origin and evolution of Triton., in: Neptune and Triton. pp. 807–877.

Ohno, K., Zhang, X., Tazaki, R., Okuzumi, S., 2021. Haze formation on triton 912, 37. https://doi.org/10.3847/1538-4357/abee82

Strobel, D.F., 2021. Atmospheric escape, in: Stern, S.A., Moore, J.M., Grundy, W.M., Young, L.A., Binzel, R.P. (Eds.), The Pluto System after New Horizons. pp. 363–377. https://doi.org/10.2458/azu_uapress_9780816540945-ch015

Strobel, D.F., Summers, M.E., 1995. Triton's upper atmosphere and ionosphere., in: Neptune and Triton. pp. 1107–1148.

Summers, M.E., Young, L.A., Gladstone, G.R., Person, M.J., 2021. Composition and structure of pluto's atmosphere, in: Stern, S.A., Moore, J.M., Grundy, W.M., Young, L.A., Binzel, R.P. (Eds.), The Pluto System after New Horizons. pp. 257–278. https://doi.org/10.2458/azu_uapress_9780816540945-ch011.

Bertrand, T., Lellouch, E., Holler, B.J., Young, L.A., Schmitt, B., Oliveira, J.M., Sicardy, B., Forget, F., Grundy, W.M., Merlin, F. and Vangvichith, M., 2022. Volatile transport modeling on Triton with new observational constraints. Icarus, 373, p.114764.

Tomasko, M.G., West, R.A., 2010. Aerosols in titan's atmosphere, in: Brown, R.H., Lebreton, J.-P., Waite, J.H. (Eds.), Titan from Cassini-Huygens. p. 297. https://doi.org/10.1007/978-1-4020-9215-2_12

Yelle, R.V., Lunine, J.I., Pollack, J.B., Brown, R.H., 1995. Lower atmospheric structure and surface-atmosphere interactions on Triton., in: Neptune and Triton. pp. 1031–1105.

Yelle, R.V., Snowden, D.S., Müller-Wodarg, I.C.F., 2014. Titan's upper atmosphere: thermal structure, dynamics, and energetics, in: Titan. p. 322. https://doi.org/10.1017/CBO9780511667398.012

Young, L.A., Bertrand, T., Trafton, L.M., Forget, F., Earle, A.M., Sicardy, B., 2021. Pluto's volatile and climate cycles on short and long timescales, in: Stern, S.A., Moore, J.M., Grundy, W.M., Young, L.A., Binzel, R.P. (Eds.), The Pluto System after New Horizons. pp. 321–361. https://doi.org/10.2458/azu_uapress_9780816540945-ch014